\documentclass[10pt,journal,compsoc]{IEEEtran}
\usepackage[caption=false,font=footnotesize]{subfig}
\usepackage{graphicx}
\usepackage{ragged2e}
\usepackage{graphics}
\usepackage{url}
\usepackage{bm}
\usepackage{multirow}

\usepackage{array}
\usepackage{rotating}
\usepackage{algorithm}
 \usepackage {algpseudocode}
\usepackage{amsmath}
\usepackage{amsthm}
\usepackage{color}

\ifCLASSOPTIONcompsoc
  \usepackage[nocompress]{cite}
\else
  \usepackage{cite}
\fi
\ifCLASSINFOpdf
\else
\fi
\begin{document}
%
\title{GenPerm: A Unified Method for Detecting  Non-overlapping and Overlapping Communities\\{\color{blue} Accepted  in IEEE TKDE}}

\author{Tanmoy Chakraborty$^{*}$, Suhansanu Kumar, Niloy Ganguly, Animesh Mukherjee, Sanjukta Bhowmick
\IEEEcompsocitemizethanks{\IEEEcompsocthanksitem TC is with University of Maryland, College Park, USA. NG and AM are with Indian Institute of Technology Kharagpur, India. SK is with University of Illinois at Urbana-Champaign, USA. SB is with University of
Nebraska, Omaha, Nebraska, USA.\newline
E-mail: tanchak@umiacs.umd.edu}}

\markboth{Journal of \LaTeX\ Class Files,~Vol.~XX, No.~XX, September~2016}%
{Shell \MakeLowercase{\textit{et al.}}: Bare Advanced Demo of IEEEtran.cls for Journals}

\IEEEtitleabstractindextext{%
\begin{abstract}
\justify
Detection of non-overlapping and overlapping communities are essentially the same problem. However, current algorithms focus either on
finding overlapping or non-overlapping communities.  We present a generalized framework that can identify both non-overlapping and
overlapping communities, without any prior input about the network or its community distribution. To do so, we  introduce a vertex-based
metric, {\em GenPerm}, that quantifies by how much a vertex belongs to each of its constituent communities. Our community detection
algorithm is based on maximizing the GenPerm over all the vertices in the network. We demonstrate, through
 experiments over synthetic and real-world networks, that GenPerm is more effective than other metrics in evaluating community structure.
 Further, we show that  due to its vertex-centric property, GenPerm can be used to unfold several inferences beyond community detection, such as core-periphery analysis and message spreading. Our algorithm for maximizing GenPerm outperforms six state-of-the-art algorithms in accurately predicting the ground-truth labels. Finally, we discuss the problem of resolution limit in overlapping communities and demonstrate that maximizing GenPerm can mitigate this problem.

\end{abstract}

}

\maketitle

\IEEEdisplaynontitleabstractindextext

\IEEEpeerreviewmaketitle

\ifCLASSOPTIONcompsoc
\IEEEraisesectionheading{\section{Introduction}\label{sec:introduction}}
\else
\section{Introduction}
\label{sec:introduction}
\fi

\IEEEPARstart{C}{ommunity} detection is one of the intensely studied problems in network science.  In general terms, a community is a set of
vertices that have more internal connections (connections to vertices within the community), than external connections (connections to
vertices outside the community). A real life analogy would be groups of people with similar interests, such as in music or in sports. This
description assumes a strict boundary where a vertex can belong to only one community. However, as often occurs in real world, a person
might be interested in both music and sports. When vertices can belong to more than one community, then the boundaries of communities
overlap. These communities are called overlapping communities as opposed to the previously mentioned non-overlapping communities. 

From the description, it is easy to see that non-overlapping communities generalize to overlapping communities. However, the existing
state-of-the-art algorithms  for detecting overlapping communities are so tuned to find overlaps that they output  overlapping regions even
when the communities are non-overlapping. On the other hand, most algorithms for finding non-overlapping communities cannot easily be
extended to their overlapping counterparts. Therefore, even though they essentially have the same objective, detecting overlapping and 
non-overlapping communities are treated as very different problems. 

 We posit that this illusory division between non-overlapping and overlapping communities is only due to the limitations of the algorithms. 
Most of the current algorithms  evaluate the communities {\em as a whole} and do not take into account the properties of the {\em individual
vertices}. There is an implicit  assumption that a vertex completely belongs to its community, and in the overlapping case, belongs {\em equally}
to all its constituent communities. This premise might not be true for all the cases. A person can be interested in both music and sports,
but she may like music slightly more than sports or vice-versa. A vertex-centric evaluation, that focuses on an individual's relation to the
community, would be able to model this non-uniformity in belonging more realistically.

Recently, we proposed a vertex-centric metric, called ``permanence'' for non-overlapping communities~\cite{chakraborty_kdd}. The
permanence of a vertex ranges from 1 (completely belonging to its community) to -1 (assigned to a completely wrong community) and provides a
quantitative measure of how much the vertex belongs  to its community. It was also theoretically proved that identifying communities
by maximizing permanence reduces the problem of resolution limit. There was however a caveat to this proof, namely that the vertices should be  tightly connected to at most one of its neighboring communities. In other words, resolution limit is mitigated only when the communities in the network are clearly separated from each other.
When a vertex becomes more tightly connected to more than one neighboring community, the communities might become overlapped. In such case,
permanence fails to handle the resolution limit.\\

\noindent{\bf Our contributions.} This observation motivated us to generalize the concept of  permanence such that it is applicable to both
overlapping and non-overlapping communities.  We develop a vertex-centric metric called ``Generalized Permanence'' (abbreviated as {\em
GenPerm}) (Section \ref{metrics}). GenPerm provides a quantitative value of how much a vertex belongs to each of its constituent
communities. Unlike permanence which does not consider the membership of the adjacent edges of a node, GenPerm penalizes the membership value of the node if its adjacent edges share multiple community memberships. This consideration leads to a more generalized version of the metric -- although GenPerm is designed for overlapping community analysis, it can easily be reduced to permanence if the underlying community structure is non-overlapping.  We show that GenPerm is very effective at evaluating goodness of communities  compared to other metrics  because: (i) 
community distributions with high GenPerm exhibit high correlation with the ground-truth, and (ii) GenPerm  is more sensitive to
perturbations in the network (Section~\ref{goodness}). Our main contributions are as follows:
\begin{itemize}
\item {\em Algorithm for unified community detection:} We shall present an algorithm, based on maximizing
GenPerm over all vertices, for unified community detection, i.e., the algorithm can detect both overlapping and non-overlapping communities.
In contrast to other algorithms, our method does not require prior information whether the network has overlapping or non-overlapping
communities or the expected number of  communities. Experimental results show that our method outperforms six state-of-the-art algorithms in
identifying ground-truth communities.

\item {\em Inferences drawn from GenPerm:} We shall show that GenPerm can be used in unfolding inferences beyond
community detection, including understanding the core-periphery structure of the communities and identifying good initiators for message
spreading. 

\item {\em Study of resolution limit for overlapping communities:} We shall conduct the first study of
resolution limit in overlapping communities and demonstrate that maximizing GenPerm can mitigate the resolution limit in overlapping 
communities.
\end{itemize}

The remainder of the paper is arranged as follows. In Section \ref{metrics}, we illustrate our rationale for formulating  GenPerm. In Section~\ref{dataset}, we present other metrics and algorithms  with which we shall compare GenPerm  and a test suite of networks on which we shall conduct our empirical evaluations. In Section \ref{goodness}, we show GenPerm is superior in evaluating the quality of a detected community and is appropriately sensitive under different perturbations. In Section \ref{applications}, we study how the different parameters in the GenPerm formula change with community structure. We also demonstrate how GenPerm can be used to analyze the core-periphery structure and select good initiator vertices for message spreading. In Section~\ref{algorithm}, we provide a heuristic to maximize GenPerm for overlapping community detection and show how this algorithm be used for detecting non-overlapping communities depending on the network under consideration. We also show how our method is more stable 
under vertex orderings and outputs more accurate communities. In Section \ref{discussion} we demonstrate that maximizing GenPerm can reduce the effect of resolution limit.  We provide a review of related work in this area in Section~\ref{related_work} and conclude in Section~\ref{end} with an overview of our future research plans.

\begin{figure}
\centering
\includegraphics[width=\columnwidth]{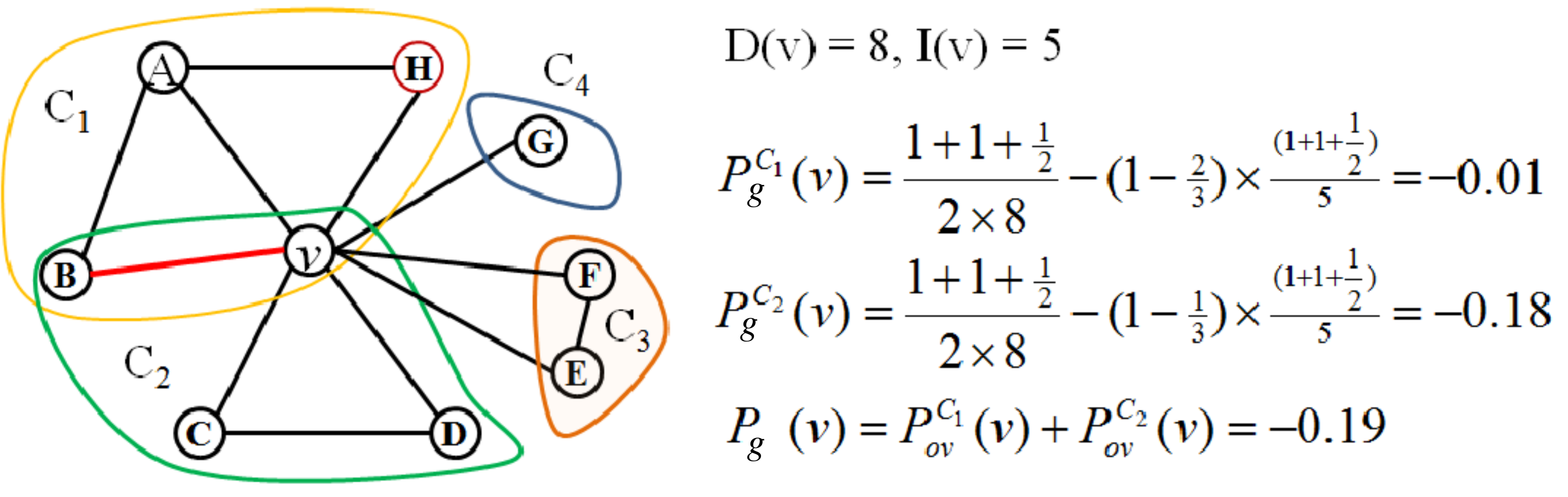}
\caption{(Color online) Toy example depicting GenPerm of a vertex $v$ which belongs to both $C_1$ and $C_2$ and has two
external neighboring communities, $C_3$ and $C_4$. The red-colored
edge shares membership in both $C_1$ and $C_2$. }\label{fig:example}
\end{figure}

\begin{table*}
\caption{Properties of the real-world networks. $N$: number of nodes, $E$: number of edges, $C$: number of
communities, $\rho$: average edge-density per community, $S$: average
size of a community, $\bar O_m$: average number of community memberships per node.}\label{datas}
\centering
 \scalebox{0.90}{
 \begin{tabular}{l||l|l|l|r|r|r|r|r|r|c}
 Networks & Node type & Edge type & Community type & N & E & C & $\rho$ & S & $\bar O_m$ & Reference\\\hline\hline
 LiveJournal & User & Friendship & User-defined group  & 3,997,962 & 34,681,189 & 310,092 & 0.536  & 40.02 & 3.09 & \cite{Leskovec} \\\hline
 Amazon & Product & Co-purchased products & Product category & 334,863 & 925,872   & 151,037 & 0.769 & 99.86  & 14.83 & \cite{Leskovec}
\\\hline 
 Youtube & User & Friendship & User-defined group & 1,134,890 &	2,987,624 & 8,385 & 0.732 & 43.88 & 2.27 & \cite{Leskovec} \\\hline
 Orkut  & User & Friendship & User-defined group & 3,072,441 & 117,185,083 & 6,288,363 & 0.245 & 34.86 & 95.93 & \cite{Leskovec}\\\hline 
 Flickr & User & Friendship & Joined group & 80,513 & 5,899,882 & 171 & 0.046 & 470.83 & 18.96 & \cite{Wang-etal12} \\\hline
 Coauthorship & Researcher & Collaborations & Research area &  391,526 & 873,775 & 8,493 & 0.231 & 393.18& 10.45 & \cite{Palla}\\\hline

 \end{tabular}}
\end{table*}

\section{Formulation of General Permanence}\label{metrics}
We first provide a brief description of the formula for permanence for non-overlapping communities ~\cite{chakraborty_kdd}, and then discuss
how we extend it for the generalized case. 

\noindent{\bf Notations and Preliminaries.} Consider a vertex $v$ with degree $D(v)$. The vertex $v$ belongs to the community $X$.
Let the number of internal connections, i.e.,  the connections
to other vertices in community $X$, of $v$ be $I(v)$\footnote{The internal connections of a node is measured in the context of
a subgraph induced by the nodes belonging to a community, i.e., $I(v) = I(v, G[C])$, where $C$ is a set of nodes and $G[C]$ is the induced
subgraph.}. Let  $c_{in}(v)$ be the internal clustering coefficient, i.e., clustering coefficient\footnote{ The clustering
coefficient for a vertex is given by the proportion of links between the vertices within its neighborhood divided by the number of links
that could possibly exist between them.}
among the neighbors of $v$ that also belong to the community $X$. The vertex $v$ might have external connections to communities other than
$X$.  Let $E_{max}(v)$ be the number of maximum connections that $v$ has to a single external community. Given these factors, permanence of
vertex $v$,  $P(v)$ is given as follows:

\begin{equation}\label{op}
P(v)=\frac{I(v)}{E_{max}(v)} \times \frac{1}{D(v)} - (1-c_{in}(v))
\end{equation}

Permanence is composed of two factors. The first factor ({\em Pull}), $\frac{I(v)}{E_{max}(v)} \times \frac{1}{D(v)}$, measures by how much
$v$ is "pulled" within its community versus how much it is pulled by the next contending community.  The novelty of this measure is that it
does not consider all the external connections, but only the connections to the community to which $v$ is likely to move. If $E_{max}(v)$ is
zero, it is set as 1.  The second factor ({\em Connectedness}), $(1-c_{in}(v))$, measures how tightly the vertex  $v$ is connected in its
own community. If the number of internal neighbors is less than two, then $c_{in}(v)$ is set to one.

\noindent{\bf Formulating GenPerm.}  To create a generalized version of permanence, we assume that the distinction between internal and
external communities still holds, and there is no overlap between any internal and external community.  However, among the internal
connections, vertex $v$ does not belong to just one community, but to a super-community, $C$, consisting of several overlapping
communities. The formulation of GenPerm concerns quantifying how much the vertex $v$ belongs to each of these overlapping communities.  

Figure~\ref{fig:example} shows an example of a vertex belonging to more than one community. From a permanence perspective, vertex $v$
belongs to the super-community consisting of vertices $A, B, C, D$ and $H$. The objective of GenPerm is to quantify by how much $v$ belongs
to the subgroups $C1$ and $C2$.

{\em Generalizing Pull.}  Let us consider the first factor in permanence, $\frac{I(v)}{E_{max}(v)} \times \frac{1}{D(v)}$. The terms
$E_{max}(v)$ and $D(v)$ remain unchanged because the possible overlaps within $C$ will not effect the maximum external connections or the
total degree of the vertex. The internal connections, however, can now be grouped into two classes: (i) the non-shared connections where an
edge belongs completely to one community, such as the edge $<v,C>$ in Figure~\ref{fig:example},  and (ii) the shared connections, such as
the edge $<v, B>$, where the edge falls in the overlap.

We posit that the internal pull a vertex experiences in a smaller community will be the sum of the non-shared edges (as in the original
permanence) plus the sum of the shared edges normalized by the number of communities over which they are shared. The normalization is
necessary so that the effect of the shared edges is counted in proportion to their commitment to the different communities. Given this, the
{\em effective internal connections} of $v$ in community $c$, where $c \in C$ is  $I^c(v)=\sum_{e\in \Gamma_v^c} \frac{1}{x_e}$, where
$\Gamma_v^c$ denotes the total set of internal edges of $v$ 
in community $c$, and $x_e$ is the number of communities in $C$ that contains the edge $e$.

{\em Generalizing Connectedness.} The second part of the formula concerns how tightly connected $v$ is with its internal neighbors.  The
permanence formulation computes this as a penalty factor, as one minus the internal clustering coefficient. The computation of the
clustering coefficient remains the same, because  we are still computing the connectedness of the neighbors. However, if the neighbors are
endpoints to a shared edge, they may not belong equally to the community. To take this into account, we normalize the internal clustering
coefficient by the fraction of the effective internal connections of the vertex $v$. Therefore the connectedness of $v$ in community $c$,
where $c \in C$ is  $(1-c_{in}^c(v))\cdot  \frac{\sum_{e\in \Gamma_v^c} \frac{1}{x_e}}{I(v)}$.

Taking these two factors together, we compute {\em GenPerm}, $P^c_{g} (v)$ of
$v$ in community $c$ as follows:
\begin{equation}
P_{g}^c(v)=\frac{\sum_{e\in \Gamma_v^c} \frac{1}{x_e}}{E_{max}(v)} \times \frac{1}{D(v)} - (1-c_{in}^c(v))\cdot \frac{\sum_{e\in \Gamma_v^c} \frac{1}{x_e}}{I(v)}
\end{equation}
This formula can be written as,
\begin{equation}\label{op}
P_{g}^c(v)=\frac{I^c(v)}{E_{max}(v)} \times \frac{1}{D(v)} - (1-c_{in}^c(v))\cdot \frac{I^c(v)}{I(v)}
\end{equation}

As shown in Equation \ref{op}, the values of GenPerm  still maintain the bounds of permanence, i.e, between 1 (vertex is completely
integrated within a clique) and -1 (vertex is wrongly assigned). The GenPerm over the network is given by the mean of the GenPerm values of
the vertices. A high GenPerm
 indicates that most of the vertices in the network are assigned to their correct communities.  Therefore, a good or accurate community
assignment is the one that maximizes GenPerm.

The total GenPerm of a vertex  over all its associated communities $C$ can be computed as $P_{g}(v)=\sum_{c\in C} P^c_{g}(v)$. 
The total GenPerm of a network $G$, with the vertex set $V$, is the average of the GenPerm values of all the vertices, i.e., 
$P_{g}(G)=\frac{1}{|V|}\sum_{v \in V} P_{g}(v)$.
Note that when $|C|=1$, i.e., the super-community consists of only one
community, $P_{g}(v)=P(v)$. These derivations show that GenPerm is indeed a generalization of permanence. It can  break down the
belongingness of a vertex to the smaller communities within the super-community. Under special boundary cases, when the super-community is
just one community, GenPerm translates to the same formula as permanence. Therefore, GenPerm is a rare metric that can be useful for both
non-overlapping and overlapping community detection.

\vspace{-5mm}
\section{Experimental SetUp}\label{dataset}
We shall now present our empirical results where we compare communities obtained through maximizing GenPerm with the ground-truth
communities from our test suite.  Since GenPerm is a generalized metric, it should be equally successful in finding both non-overlapping and
overlapping communities.  However, we observed that almost all networks have some degree of overlapping communities, even the ones claiming
to have (or create) only non-overlapping communities. The ground-truth provided by the apparent non-overlapping communities merely ignore
the overlaps that occur. In these cases, the ground-truth cannot be accurately compared with the communities produced by GenPerm, because
GenPerm identifies the hidden overlaps as well.

Therefore, for an equitable comparison  against ground-truth, we shall conduct experiments  on datasets that are known to have overlapping
communities, and use metrics for evaluating overlapping communities to assess our results. To demonstrate that maximizing GenPerm is also
successful in identifying non-overlapping communities, we show its performance on idealized networks consisting of loosely connected
cliques.

\subsection {Test Suite of Networks}\label{sample}

 Our synthetic test set consists of 
LFR\footnote{\url{http://sites.google.com/site/andrealancichinetti/files.}} benchmark networks \cite{Lancichinetti}. 
LFR provides a rich set of parameters to control the network topology, including  the percentage of overlapping nodes $O_n$, number of communities to which a node belongs $O_m$,  the number of nodes $n$, the
mixing parameter $\mu$, the average degree $\bar k$, the maximum degree $k_{max}$, the maximum community size $c_{max}$, and the minimum
community size $c_{min}$. 
Unless otherwise stated, we generate LFR graphs with the following configuration: $\mu$ = 0.2, $N$=1000, $O_m$=4, $O_n$=5\%; other parameters are set to their default
values\footnote{\url{https://sites.google.com/site/santofortunato/inthepress2}}. Note that for each parameter configuration,
we generate 100 LFR networks, and the values in all the experiments are reported by averaging the results. 

We also use six real networks whose underlying ground-truth community structures are known a priori. The properties of these networks are
summarized in Table \ref{datas}. \\

\noindent{\bf Sampling real-world networks.} Most of the baseline community detection
algorithms (Section \ref{baseline})
do not scale to large networks. Therefore, we use the technique from \cite{Leskovec} to obtain subnetworks with overlapping community structure from
 larger networks. We pick a random node $u$ in graph $G$ that belongs to at least two
communities. We then take the subnetwork to be the induced subgraph of $G$ consisting of all the nodes that share at least one ground-truth
community membership with $u$. In our experiments, we created 500 different subnetworks for each of the six real-world datasets, and the
results are averaged over these 500 samples\footnote{Note that in the rest of the paper while reporting the average results over many samples, we notice that the variance is significantly less. Therefore,  we omit reporting the variance for the sake of brevity.}. The properties of the sampled graphs are reported in SI Text.

\begin{figure*}[ht!]
\centering
\scalebox{0.4}{
\includegraphics{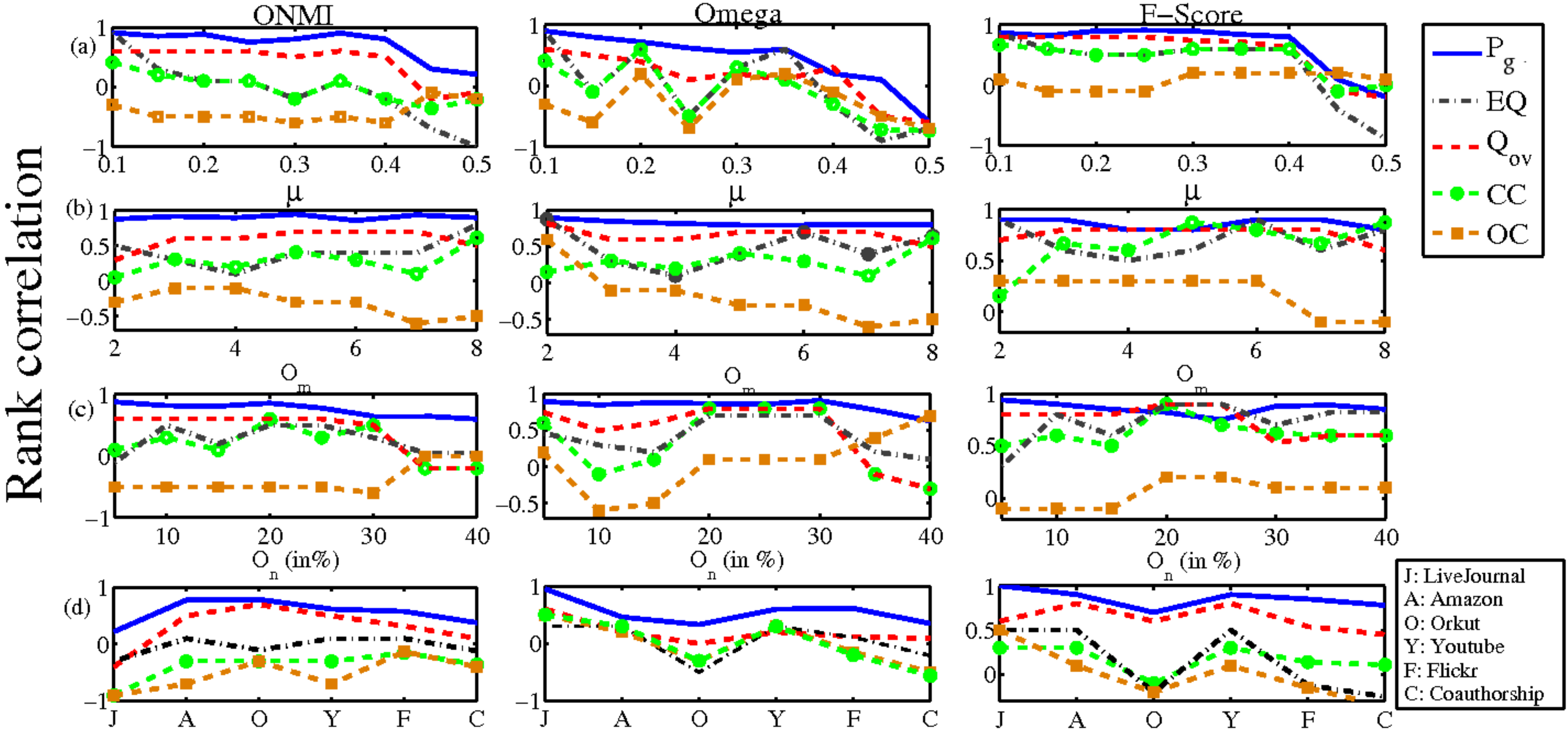}}
\caption{(Color online) Spearman's rank correlation between five scoring metrics with three validation measures for LFR
by varying (a) $\mu$ (where $N$=1000, $O_m$=4, $O_n$=5\%), (b) $O_m$ (where $N$=1000, $\mu$=0.2, $O_n$=5\%), (c) $O_n$ (where $N$=1000,
$\mu$=0.2, $O_m$=4) and (d) real-world networks. Note that the values of $O_n$ are expressed in \% of $N$. }
\label{correlation}
\end{figure*}

\subsection{Overlapping Community Scoring Metrics} \label{all_metric}
We compare GenPerm with the following metrics 
for evaluating  quality of the overlapping community structure. \\
 {\bf $\bullet$ Modularity:} Shen et al. \cite{Shen} introduced $EQ$, an adaptation of modularity function \cite{NewGir04} for overlapping
communities:

\begin{equation}
 EQ=\frac{1}{2m}\sum_{c\in C} \sum_{i\in c,j\in c} \frac{1}{O_iO_j} \left[ A_{ij} - \frac{k_ik_j}{2m} \right]
\end{equation}
where, $A_{ij}$ indicates ($i,j$) entry in the adjacency matrix $A$, $m$ is the number of edges in the graph, $C$ is the
set of communities, and $O_i$ is the number of communities to which the node $i$ belongs.

 L{\'a}z{\'a}r  et al. \cite{Vicsek}  provided a more complex formulation of modularity for overlapping community structure:
\begin{equation}
 Q_{ov}=\frac{1}{|C|} \sum_{c\in C} \Bigg [  \frac{\sum\limits_{i\in c}  \frac{\sum\limits_{j\in c, i \neq j} A_{ij} -
\sum\limits_{j \notin c} A_{ij}}{d_i\cdot s_i } }{n_{c}} \cdot  \frac{n^e_{c}}{	\dbinom{n_{c}}{2}}  \Bigg ]
\end{equation}
where $C$ is the set of communities, $n_{c}$ and $n^e_{c}$ are the number of nodes and edges that community
$c$
contains respectively, $d_i$ is the degree of node $i$, and $s_i$ is the number of communities to which $i$ belongs.

\noindent{\bf $\bullet$ Community Coverage (CC):} This metric  \cite{nature2010} counts the fraction of nodes that belong to at
least
one community of three or more nodes. A size of three is chosen since it constitutes the smallest non-trivial community structure. 

\noindent{\bf $\bullet$ Overlap Coverage (OC):} This metric  \cite{nature2010} counts the average number of node memberships in
non-trivial 
communities (of size at least three).

\subsection{Baseline Algorithms} \label{baseline}
We compare our method with the following state-of-the-art overlapping community detection algorithms. 
These algorithms together
cover the types of overlapping community detection heuristics mentioned in
\cite{Xie_2013s}:
(a) {\it   Local expansion and optimization:} OSLOM
\cite{oslom}, EAGLE \cite{Shen}
(b) {\it Agent-based dynamical algorithms:} COPRA \cite{Gregory1}, SLPA \cite{Xie}
(c) {\it Fuzzy detection using mixture model:} MOSES \cite{moses}, BIGCLAM \cite{Leskovec}.

\subsection{Community Validation Metrics} \label{validation}
 We use the following 
metrics that quantify the level of correspondence between the ground-truth communities and those obtained from an algorithm: (a) 
Overlapping Normalized Mutual Information (ONMI) \cite{McDaid}, (b) Omega ($\Omega$) Index \cite{Gregory}, (c)  F-Score \cite{Leskovec}.
The higher the value of the metrics, the closer the match with the ground-truth (see SI Text).

\section{GenPerm for Community Evaluation}\label{goodness}
We  now demonstrate that GenPerm is a better metric for evaluating the quality of communities than the metrics listed in 
Section \ref{all_metric}. 

\subsection{Correspondence to Ground-truth Structure}
We show using rank correlation (following the approach in \cite{steinhaeuser2010}) that communities with high GenPerm best correspond to the
ground-truth. To do so, we perform the following steps (Table \ref{table1}
is used to illustrate this experiment): \\
(i) For each network, we execute six algorithms from Section
\ref{baseline} and obtain the communities; (ii) For each of these communities, we compute the GenPerm value and the scoring metrics
discussed
in Section \ref{all_metric}; (iii) The algorithms are then ranked based on the value of these metrics, with the 
highest rank given to the highest value (as an example, in the second column of Table \ref{table1}, we show the values for GenPerm and the
corresponding ranks of the algorithms); (iv) The
community structures are compared with the ground-truth labels in
terms of the validation metrics, ONMI, Omega Index and F-score; (v) The algorithms are again ranked based on the values of each of the
validation metrics (highest
value/best match has the best rank) (in the third column of Table \ref{table1}, the algorithms are ranked based on the values of
ONMI); (vi) Finally, we obtain Spearman's rank correlation between 
the rankings obtained for a scoring metric (step (iii))  and each of the ground-truth validation metric (step 
(v)). 

We posit that since these two types of measures are orthogonal, and because the
validation metrics generally 
provide a stronger measure of correctness due to direct correspondence with the ground-truth structure, the
ranking of a good community scoring  metric should ``match" with those of the validation metrics. We compare the
relative ranks instead of the absolute values, because the range of the values is not commensurate 
across the quantities and therefore the rank order is a more intrinsic measure.

\begin{table}[!h]
\caption{For Amazon network, the values of  average GenPerm  of the network on the output obtained from
different algorithms and ONMI with respect to the ground-truth community. The ranks of the algorithms (using dense ranking) are shown
within parenthesis. }\label{table1}
\centering
 \scalebox{0.9}
 {
\begin{tabular}{|c||c||c|}
\hline
{\bf Algorithms} & {\bf $P_g$} & {\bf ONMI}\\\hline    \hline

OSLOM	&	0.63 (1)	&	0.73 (1)	\\\hline 
EAGLE	&	0.53 (4)	&	0.52 (4)	\\\hline 
COPRA	&	0.60 (2)	&	0.70 (3)	\\\hline 
SLPA	&	0.56 (3)	&	0.70 (3)	\\\hline 
MOSES	&	0.41 (5)	&	0.50 (5)	\\\hline 
BIGCLAM 	&	0.60 (2)	&	0.71 (2)	\\\hline 
\end{tabular}
}
\end{table}

 Figure \ref{correlation} shows the correlation values for different LFR networks (where $\mu$, $O_m$ and $O_n$ are varied) and six real-world
networks (the values are reported by averaging over 500 subnetworks in each case). The vertical panel corresponds to a validation measure. Each line in a panel corresponds to
an evaluation function. For all the cases, the lines corresponding to $P_{g}$ (average GenPerm  of the network) is higher than the
other metrics, which is mostly followed by $Q_{ov}$, $EQ$, $CC$ and $OC$. Therefore, we conclude that GenPerm can evaluate communities better than other metrics. We also notice that with the increase in $\mu$, the line corresponding to GenPerm tends to decrease due to the deterioration of the ground-truth community structure. However, the GenPerm-based rank correlation is almost consistent with the increase of $O_m$ and $O_n$, which is not as consistent for the other metrics. This indeed indicates that GenPerm does not get affected much with the extent of overlap among communities as long as the explicit community structure is retained in the network.

 \begin{figure*}[ht!]
\centering
\scalebox{0.3}{
\includegraphics{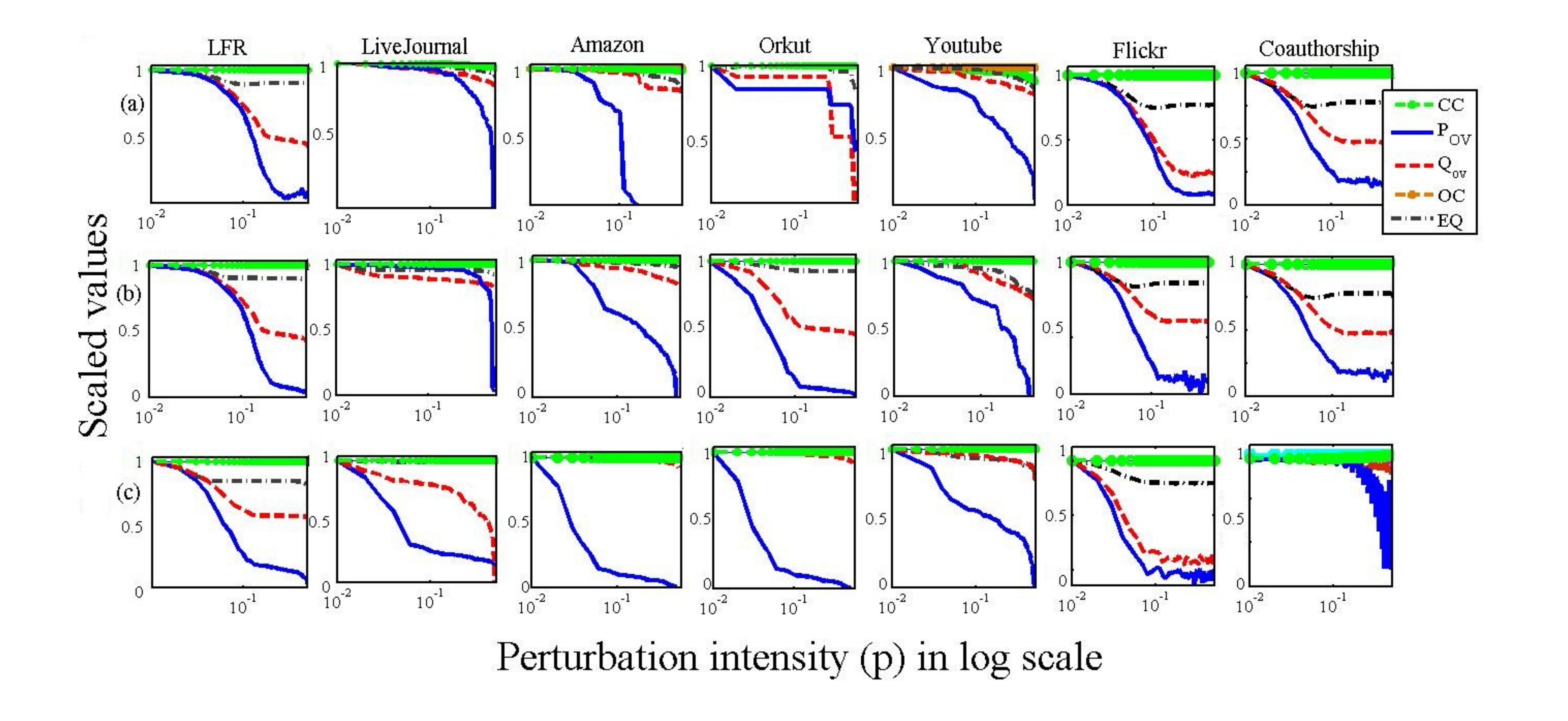}}
\caption{(Color online) Change in the value of five overlapping community scoring functions with the increase of perturbation
intensity ($p$) in three perturbation strategies: edge-based (top panel), random (middle panel) and community-based (bottom panel)
strategies for a LFR and six real-world networks. Each data point is the average over 500 simulations. The values of each function are normalized by the maximum value obtained from that function. Most of the cases, the lines for community coverage ($CC$) and overlap coverage ($OC$) are starkly juxtapose because of their high similarity.}\label{fig:perturbation}
\end{figure*}

 \subsection{Robustness to Perturbations}\label{sensitivity}
So far we have examined the ability of different scoring metrics to rank algorithms according to their goodness. In this section,
 we evaluate the metrics on their robustness under perturbations to the network. A metric is robust  if its value 
is commensurate with the change in ground-truth communities. This means that 
under
small perturbations 
to the ground-truth,  the results change slightly.  However, if the ground-truth labels are highly perturbed such that the
underlying community structure gets largely deformed, 
then a good community scoring metric should produce a low score. Given a graph $G$=$<${\it V},{\it E}$>$  and  \emph{perturbation
intensity}  $p$, we  start with the ground-truth community $S$ and  modify it by using the following
strategies (as used in ~\cite{chakraborty_kdd}). 

 {\bf (i) Edge-based} perturbation selects an inter-community edge $(u,v)$ where $u \in S$ and $v \in S'$ ($S \neq S'$) and 
assigns $u$ to $S'$ and $v$  to $S$. This continues for $p \cdot |E|$ iterations.

{\bf (ii) Random} perturbation selects two random nodes $u \in S$ and $v
\in S' $ ($S \neq S'$) and then swaps their memberships. This  continues for $p \cdot |V|$ iterations. 

{\bf (iii) Community-based} perturbation takes each community $s$ from the ground-truth structure $S$  and swaps $p \cdot |s|$ constituent nodes in $s$  with non-constituent nodes.

 We perturb different networks using these three  strategies for values of $p$ ranging between 0.01 to 0.5,  and
compute five community scoring metrics, i.e., $P_{g}$, $EQ$, $Q_{ov}$, $CC$, $CC$. Figure~\ref{fig:perturbation}
shows the representative results only for one LFR ($\mu$ = 0.2, $N$=1000, $O_m$=4, $O_n$=5\%) and all real-world networks. For all three
strategies, the value of the
scoring
metrics tends to decrease with the increase in $p$; the effect is most pronounced in community-based strategy. For each network, once $p$ has reached a certain threshold, the decrease is much
faster in GenPerm.
This happens because the internal structure of a community completely breaks down if the perturbation is taken beyond a point and thus has an avalanche effect on the value of the clustering coefficient ($c_{in}^c(v)$ in Equation
\ref{op}).

 \vspace{-4mm}
\section{Inferences drawn from GenPerm}\label{applications}

We show how, due to its vertex-centric view, GenPerm can help us understand the distribution of nodes in the constituent communities.
 We also show how ranking vertices in increasing order of GenPerm can be used to explore the core-periphery structure of a community, and in selecting initiator nodes during message spreading in networks. Finally, we show how the layered structure obtained from GenPerm can affect the performance of the community detection algorithms.

\subsection{Understanding  the Community Structure }  \label{permval}

\begin{figure}[!h]
\centering
\scalebox{0.9}{

\includegraphics[width=\columnwidth]{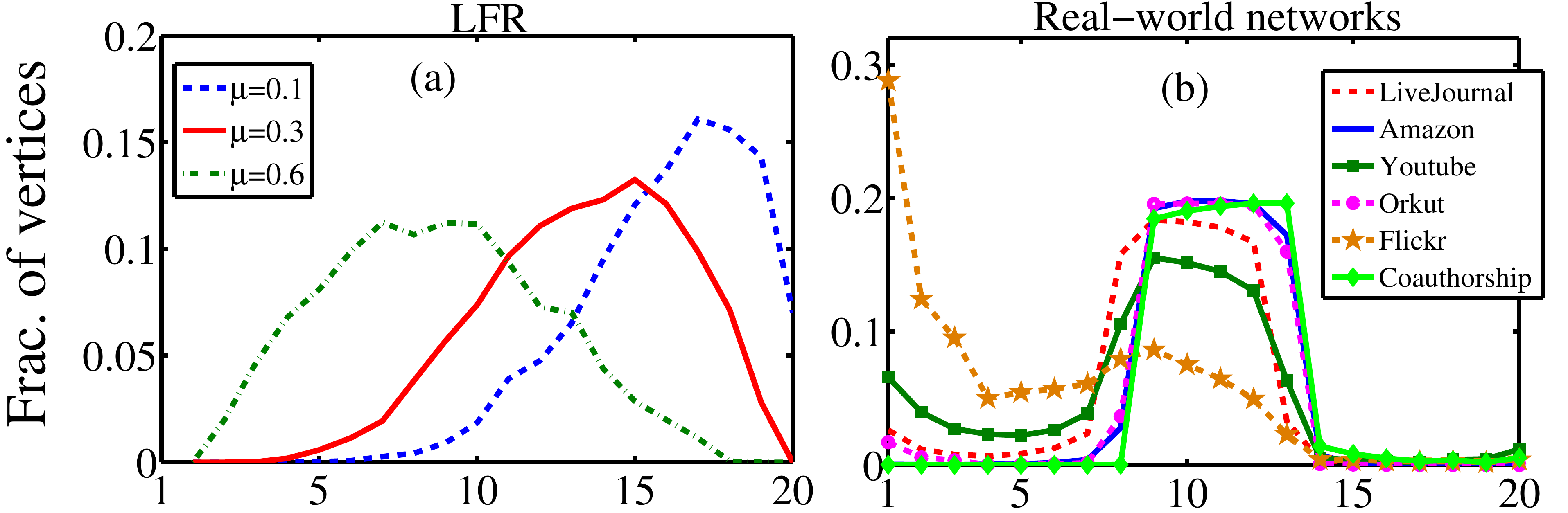}
}
\scalebox{0.9}{

\includegraphics[width=\columnwidth]{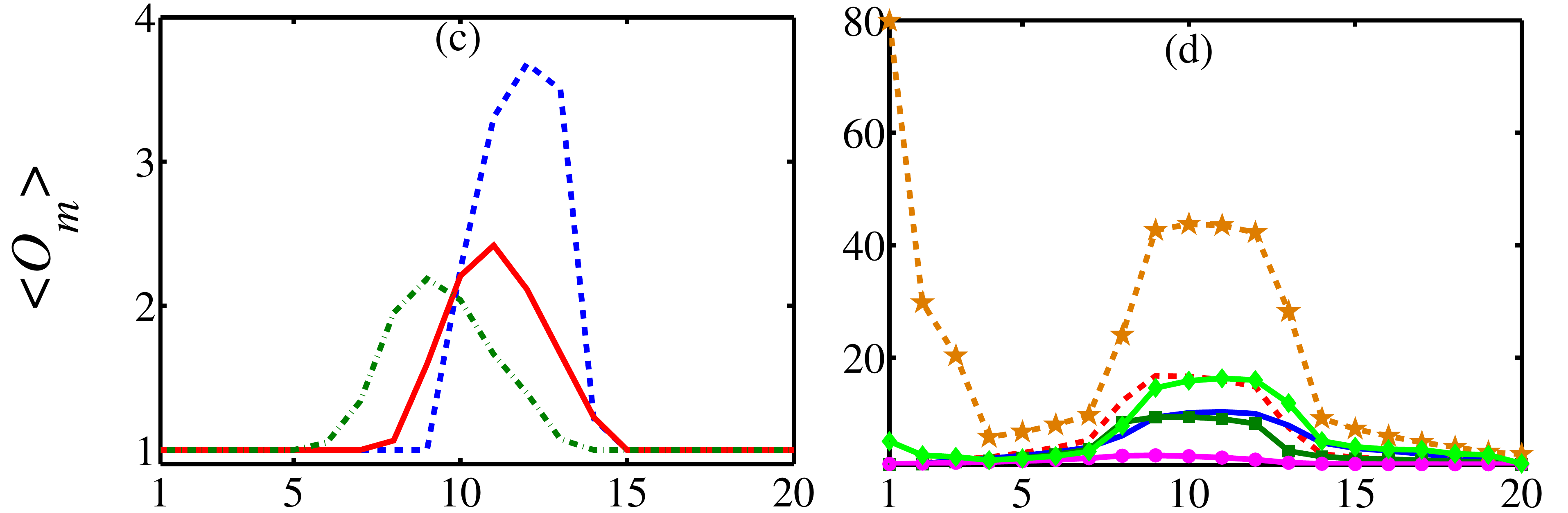}
}
\scalebox{0.9}{

\includegraphics[width=\columnwidth]{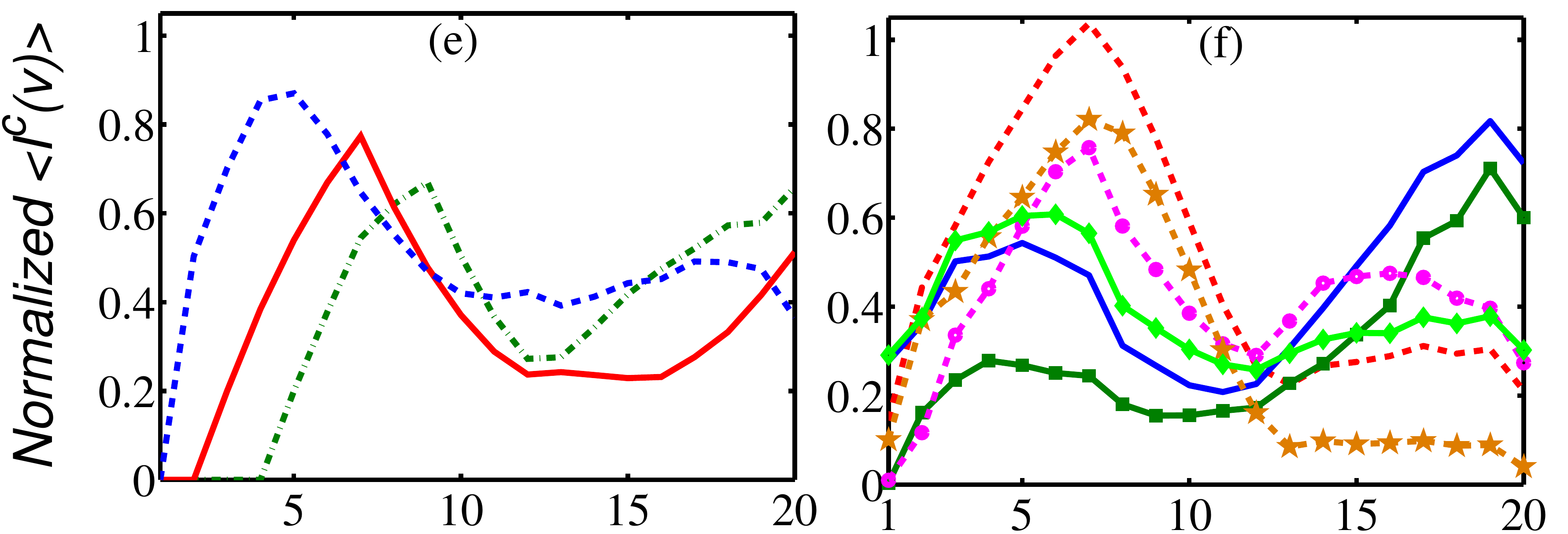}
}
\scalebox{0.9}{

\includegraphics[width=\columnwidth]{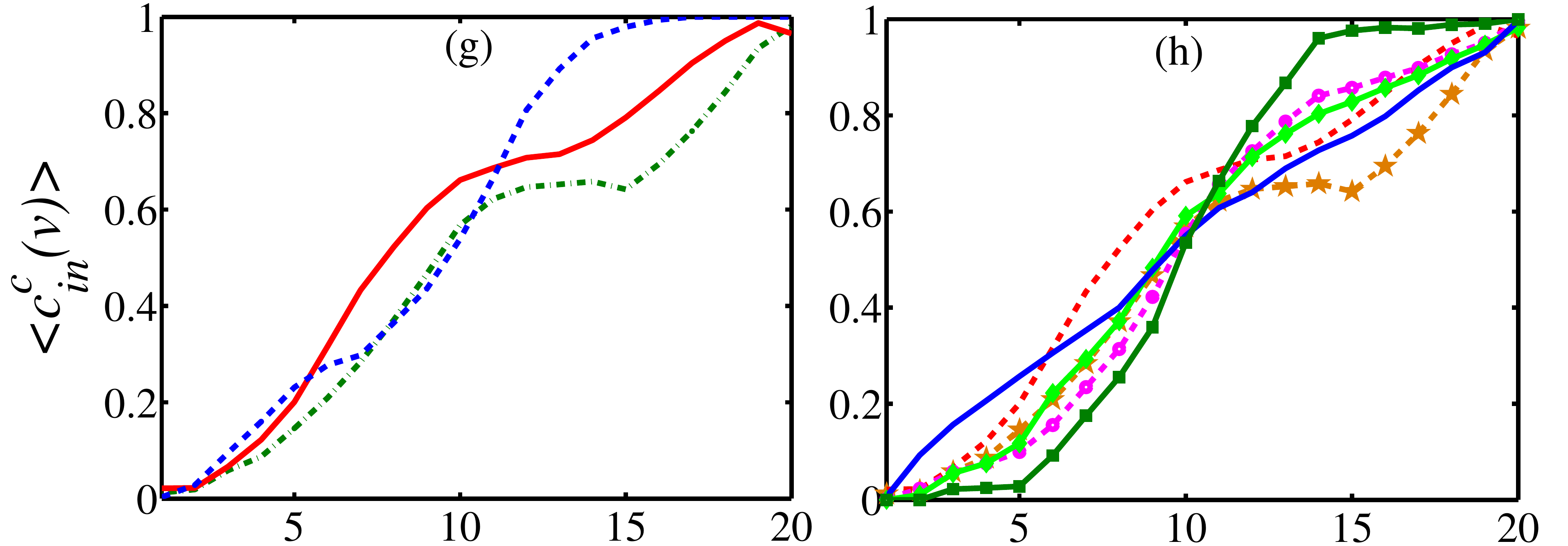}
}
\scalebox{0.9}{

\includegraphics[width=\columnwidth]{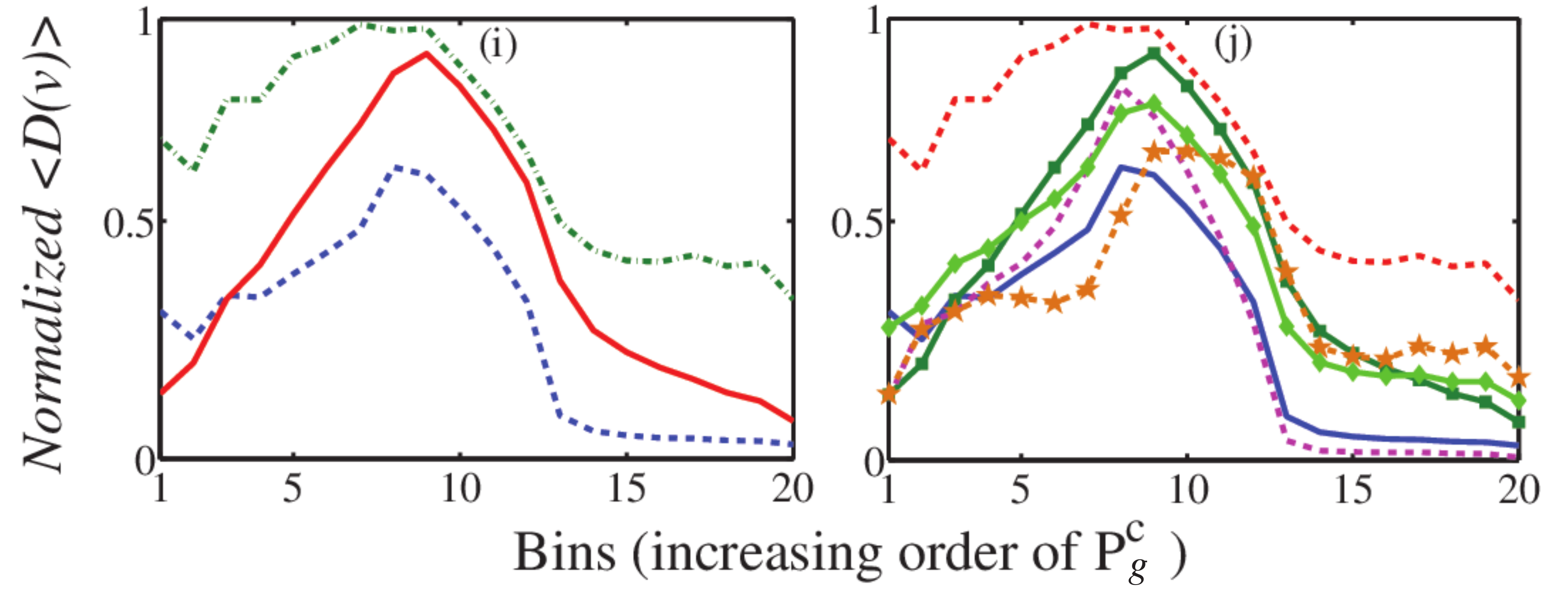}
}
\caption{(Color online) The relation of average $P_{g}^c$ with (a)-(b) fraction of vertices, (c)-(d) $<O_m>$, average community memberships
per node, (e)-(f) $<I^c(v)>$, average internal degree  (normalized by the maximum value), (g)-(h) $<c_{in}^c(v)>$, average internal
clustering coefficient, (i)-(j) $<D(v)>$, average degree of nodes  for LFR and real-world
networks. The value of $P_{g}^c$ of vertices in each community is equally divided into 20 buckets indicated in
x-axis (bin 1: $-1\leq P_{g}^c<-0.9$, ..., bin 20: $0.9\leq P_{g}^c\leq1$).}\label{fig:perm_dist}
\end{figure}

We study the distribution of GenPerm for each node-community pair. We compute $P_{g}^c$ of each vertex on the ground-truth communities of the benchmark
networks. In Figure \ref{fig:perm_dist}, we divide the values of $P_{g}^c$ ranging from -1 to 1 into 20 bins on x-axis where the low (high) numbered bins contain nodes
with lower (higher) $P_{g}^c$. On the y-axis we plot different features as described below.

\noindent{\bf Fraction of vertices.}  In Figure \ref{fig:perm_dist}(a) - \ref{fig:perm_dist}(b), we plot in y-axis the fraction of vertices
present in the network. 
This curve follows a Gaussian-like distribution, i.e., there are few vertices with very high or very low $P_{g}^c$ values, and the majority
have intermediate values. In Figure \ref{fig:perm_dist}(a), the peak  shifts from left to right with the decrease of $\mu$  (other
parameters of LFR are constant).  The shift in the peak shows that as the structure of the communities gets more
well-defined, most vertices move towards higher values of GenPerm. 
The real-world networks, except Flickr  show a similar Gaussian distribution in Figure \ref{fig:perm_dist}(b), where most of the
vertices fall in medium $P_{g}^c$ range. 
For Flickr
network, we notice in Table \ref{datas} that the communities are large  (high $S$) and their edge density is low (low $\rho$). Moreover,
 most of the vertices in Flickr have low internal clustering
coefficient (0.12, where the average for the other networks is 0.31). These factors contribute to the very low $P_{g}^c$. 

\noindent{\bf Number of constituent communities.}
We plot the number of communities to which a vertex belongs in Figure \ref{fig:perm_dist}(c) - \ref{fig:perm_dist}(d). This pattern also follows a Gaussian distribution for LFR networks. This
indicates that vertices exhibiting average $P_{g}^c$ tend to belong to multiple communities, than vertices with very high or low values. 
This is because vertices with very high $P_{g}^c$ values, such as 1, tend to be tightly integrated within one community and those with low
$P_{g}^c$ values have low degree and therefore are not part of many communities.

\noindent{\bf Parameters of GenPerm. } In Figure \ref{fig:perm_dist}(e) - \ref{fig:perm_dist}(f) we plot the average value of the effective
internal connections, $I^{c}(v)$, of vertices for
each $P_{g}^c$ bin. For each network, the values of $I^{c}(v)$ are normalized by the maximum value. We observe the
minimum value of $I^{c}(v)$ is in the middle bins. This is because the vertices participating in many communities contribute a 
small fraction of internal edges to each community. Next, we plot the  average $c_{in}^c(v)$ of the vertices in each bin. Here we notice  a nearly linear relation. Finally, we plot the degree  in Figure \ref{fig:perm_dist}(i) - \ref{fig:perm_dist}(j). We observe that vertices with
higher degree exhibit medium $P_{g}^c$. This is because vertices with high degree, but low effective internal connections, belong to multiple communities. 

Taking these results together, we conclude that the effective internal connections and the degree are good indicators of the number of communities to which the vertex belongs and the internal clustering coefficient is a good indicator of the GenPerm value of the vertex.

\begin{figure}[!h]
\centering
\scalebox{0.22}{
\includegraphics{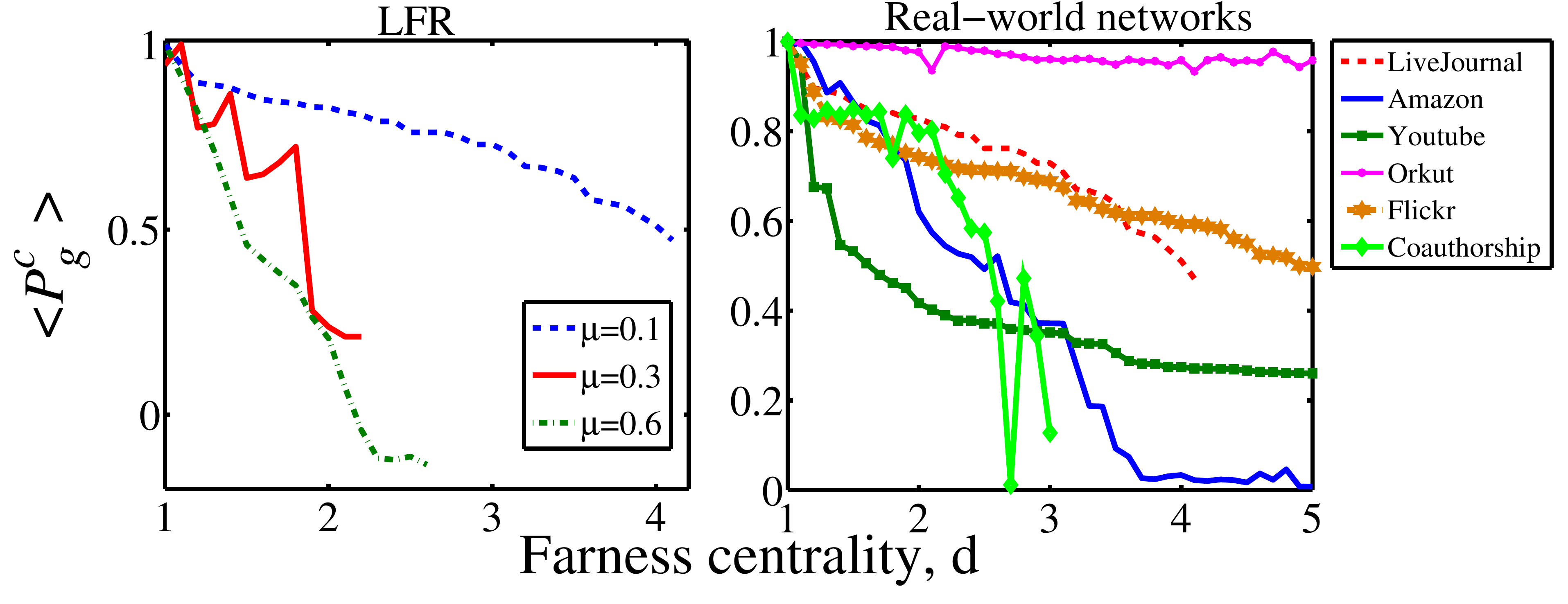}}
\caption{(Color online) Community-wise average GenPerm, $<P_{g}^c>$ of vertices as a function of farness centrality $d$ for
LFR and real-world networks.}
\label{fig:farness}
\end{figure}

\vspace{-3mm}
\subsection{Core-periphery Structure of Community} \label{core} 
To study the relation between GenPerm and the core-periphery of the communities,
 we use {\em farness centrality} ($d$) \cite{Yang14} as a measure of the position of a vertex within a community. 
To measure farness centrality 
for each community, we construct the induced subgraph of all the nodes in the community and
measure average shortest path for each vertex within this subgraph\footnote{{\scriptsize Farness centrality is just the opposite of closeness centrality in a connected component.}}. The lower the value of $d$ for a vertex, the closer the vertex is to
the core of the community.  
We plot average $P_{g}^c$ of vertices as a function of $d$ in Figure \ref{fig:farness}. We observe that for both LFR and
real-world networks, average $P_{g}^c$ decreases with the distance from the center of the community. 
Therefore,  the value of GenPerm is a strong indicator of the position of the vertex in the  community. We observe that in coauthorship network, at $d=2.8$, the value of GenPerm drops suddenly, the reason being very less internal clustering coefficient of the nodes at that region of the community.

 We next measure  assortativity ($r$)\footnote{{\scriptsize Assortativity ($r$)
lies between -1 and 1. When $r$ = 1, the network is said to have perfect
assortative patterns, when $r$ = 0 the network is non-assortative, while at $r$ = -1 the network is completely disassortative.}}
 \cite{Newman-assort-2003} with respect to GenPerm to evaluate the extent to which the vertices in a community $c$ to attach to other vertices with similar GenPerm. For each ground-truth communities, we calculate $P_{g}^c$ of all the vertices. Then we divide the values of $P_{g}^c$ into
20 bins. Vertices falling in the same bin are assumed to have similar  $P_{g}^c$, and then we
measure the assortativity, $r$, for the vertices in the community $c$. We compare this measure with the
degree-based assortativity of vertices in each community. 
We observe, in  Table \ref{assor}, that  both the synthetic and the real-world networks are highly assortative in terms of GenPerm, i.e., nodes exhibiting same $P_{g}^c$ value are often connected with each other than the nodes with different $P_{g}^c$ values. This result presents a new perspective about the configuration of nodes within a community: {\em vertices in a community  tend to be highly connected  within each layer than across layers. Moreover communities are  organized into several layers, 
with each layer being composed of vertices with similar GenPerm.}

\begin{table}[h!]
\caption{Average of the assortativity scores, $<r>$ (degree-based and $P_{g}^c$-based) of the communities per network.}\label{assor}
\centering
\scalebox{0.85}{
 \begin{tabular}{|c|c|c|c|}
\hline
$<r>$ & LFR ($\mu=0.1$) & LFR ($\mu=0.3$) & LFR ($\mu=0.6$) \\\hline
Degree-based  &   -0.045              &      -0.018           &      0.139            \\\hline
$P_{g}^c$-based &    0.645          &    0.483             &   0.421 \\\hline
 \end{tabular}}

\scalebox{0.75}{
 \begin{tabular}{|c|c|c|c|c|c|c|}
\hline
$<r>$         & LiveJournal & Amazon & Youtube & Orkut & Flickr & Coauthorship \\\hline
Degree-based  &    0.037         &  -0.275      &  -0.182       &  0.221     &   -0.098     &  0.281         \\\hline
$P_{g}^c$-based &  0.465        &  0.497      &   0.438       &   0.528    &   0.402     &  0.469 \\\hline
 \end{tabular}}

\end{table}

\subsection{Effect on Community Detection Algorithms}
\begin{figure}[!h]
\centering
 \includegraphics[width=\columnwidth]{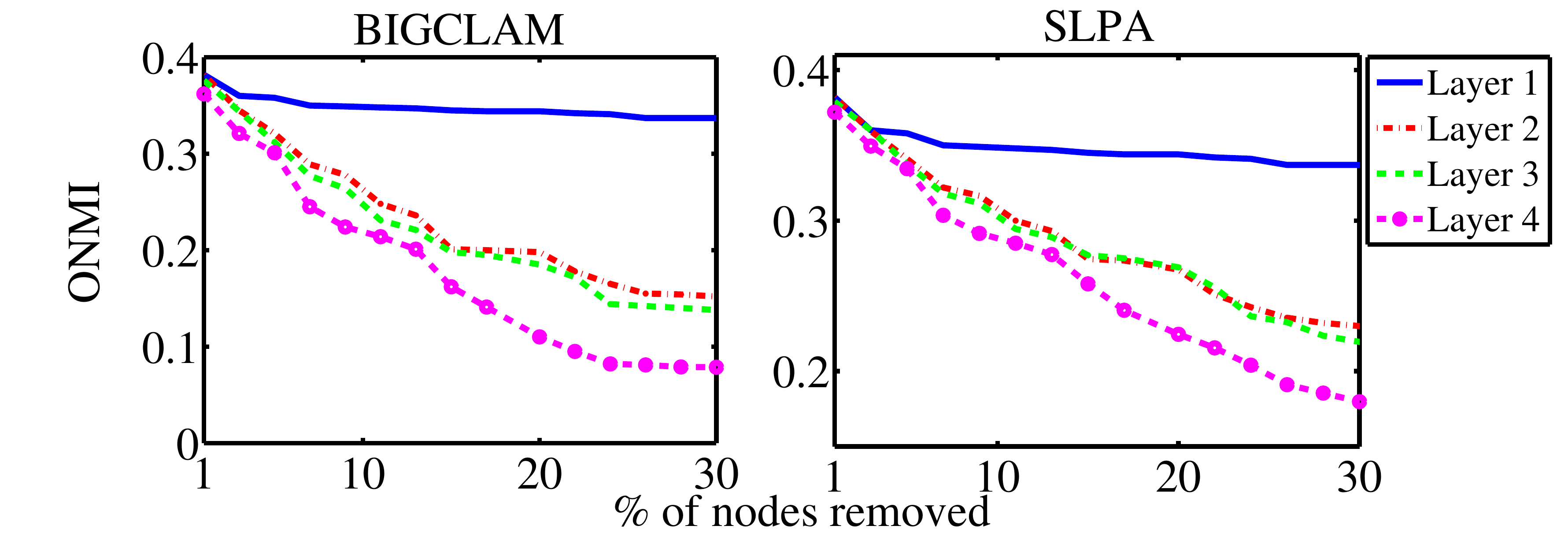}
 \caption{(Color online) Change in the value of ONMI with the removal of nodes from different layers within the communities for Amazon
network. Layer 1 corresponds to the outer-most layer, and layer 4 correspondences to the inner-most layer. Each point in the line is an
average of 100 simulations.}\label{deteo}
\end{figure}

In Section \ref{core}, we have noticed that GenPerm provides a gradation/ranking of nodes in each community that in turn produces a layered structure inside the community. We believe that this layered structure
also affects the performance of different community finding algorithms.  We consider the following two overlapping community detection
algorithms -- BIGCLAM \cite{Leskovec} and SLPA \cite{Xie}, and see how these algorithms get affected after removing selected nodes (based on the values of GenPerm) from the network based. To begin with, we consider the ground-truth community structure of a network. Then for each community, we calculate the GenPerm for all the constituent vertices and  divide the range of GenPerm into four equal bins, representing four  layers.  Following, we remove $x\%$ of nodes (refer to as set
$R$) from
each layer of the community in isolation ($x$ varies from 1 to 30). At a time, we remove nodes from one layer in all the communities, run the two overlapping community finding algorithms and see for each algorithm, how the community structure obtained from the resultant
graph (refer to as community $C_x$) matches with the
community structure obtained initially without removing any nodes (refer to as community $C_1$). We measure the correspondence between two
community structures using
ONMI. Note that for fair comparison, we first remove the set of nodes $R$ from the
initial community structure $C_1$ and then compare two communities $C_1$ and $C_x$. For each value of $x$, the simulations are done 100 times and the average value is reported. Figure \ref{deteo} shows the change in the value of ONMI with the increase in the percentage of removed nodes, $x$ from each layer for Amazon network (the
results are identical for other networks). We observe that the rate of decrease in ONMI tends to increase when the nodes are removed from
inner layers. This result indicates that the nodes in the inner layers are mostly responsible to form the community structure. The removal
of nodes from the core of a community results in producing amorphous community structure. We believe that this information is quite helpful
in the context of community-centric immunization \cite{PhysRevE.65.036104} and targeted attacks \cite{albert2000error} in the network.

\vspace{-5mm}

\subsection{Initiator Selection for Message Spreading}\label{sec:spreading}

In message spreading \cite{ChierichettiLP10}, a set of source vertices (initiators) start sending
a message. 
At every time step, a vertex containing  the  message transfers the message to one of its neighbors who does not  have the message.
The algorithm terminates when all vertices have received the message.  The selection of the initiators is critical to how quickly the message spreads.

\begin{figure}[!h]
\centering
\scalebox{.85} {
\includegraphics[width=\columnwidth]{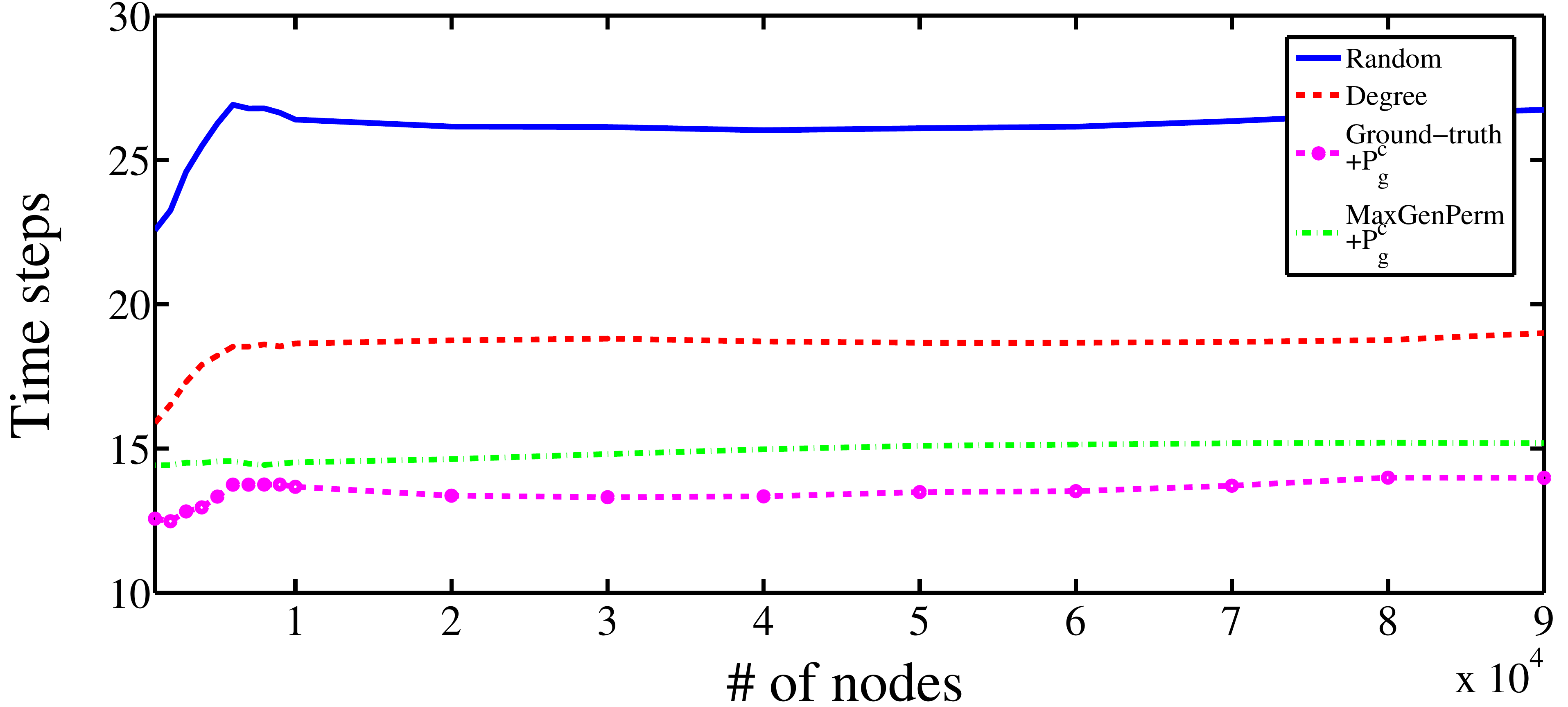}
}
\caption{(Color online) Number of time steps required to spread a message in LFR network by varying the number of nodes.}\label{spreading}
\label{farness}
\end{figure}

A fundamental issue in message spreading is the selection of initiators. 
The traditional practice is to select initiators based on the
degree of nodes, which was proved to be more useful than the random node selection in terms of average time steps required to broadcast the
message \cite{Demers:1987}. 
We posit that initiator selection based on  $P_{g}^c$ would  help in disseminating the
message more quickly. For our experiments, we create  LFR network with the number of nodes from 1,000 to 90,000, keeping the other parameters constant. We  select initiators based on the following criteria separately: (i) random, (ii) highest degree, (iii) highest $P_{g}^c$ as per ground-truth communities and (iv) highest $P_{g}^c$ based on communities obtained by our proposed algorithm, MaxGenPerm (described in Section \ref{algorithm}).  For each network configuration, we run 500 simulations. Figure \ref{spreading} shows the average number of time steps required for the message to reach all the vertices. We observe that $P_{g}^c$-based initiator selection from ground-truth communities requires minimum time steps and 
$P_{g}^c$-based initiator selection for the communities obtained from MaxGenPerm is a close second. 
These results thus highlight 
the importance of $P_{g}^c$-based ranking within a community.

\vspace{-3mm}

\section{Community Detection by Maximizing GenPerm}\label{algorithm}
 
We develop MaxGenPerm (pseudocode in Algorithm 1), an iterative method for maximizing GenPerm (the objective function of the algorithm). Such iterative methods have shown to be very
effective in optimizing  metrics for finding non-overlapping communities such as modularity~\cite{blondel2008} and
permanence~\cite{chakraborty_kdd}. Here the challenge is to update the values based on one or more  communities. 

 {\bf Initialization:} The algorithm starts with initializing each edge of a network as a separate community.  Thus a vertex can belong to more than one community.
 
{\bf Update:} For each iteration, GenPerm of vertex $v$, $P^c_{g}(v)$, in each of its neighboring communities $c$ is computed. If the value of $P^c_{g}(v)$ is greater than zero, vertex $v$ is assigned to the community $c$. This creates a new set of communities, $TempComm$, to which $v$ belongs. If the total GenPerm of the new community set, $TempComm$, is greater than the previous community set $CurComm$, to which $v$ belonged then the community set of $v$ is updated. In this manner, the communities of $v$ are updated per iteration. In this process, one might encounter singleton communities (communities with single node); we retain the singleton communities without merging them further. The algorithm ends when there is no further improvement in the total GenPerm for all the vertices or the maximum number of iterations  is
reached. 

{\bf Convergence:} There is no theoretical guarantee of the convergence of the algorithm. However we noticed that for both synthetic and large (complete) real-world networks, the algorithm converges before reaching the maximum number of iterations (we set it to 15) as follows: LFR ($n$=10,000, $\mu$=0.2, $O_n$=5\%, $O_m$=4): 4, LiveJournal: 12,  Amazon: 6,  Orkut: 10, Youtube: 9, Flickr: 5, and Coauthorship: 6.

\noindent{\bf Complexity analysis.} Let a vertex $v$ has degree $d_v$ and $C^i$ number of neighboring communities at iteration $i$;  out of
these $C_v^i$ are communities to which $v$ belongs. 

At each iteration for vertex $v$,  we add its GenPerm over all $C_v^i$ number of communities.  Let this time be $T.C_v^i$.
At each iteration for vertex $v$,  we go through all the communities $C^i$, to compute the update of GenPerm. Let the internal degree of
$v$ in a community $c \in C_v^i$ be $I_c$. The time to compute GenPerm in $c$ is dominated by computational time of the clustering coefficient. It is
therefore $O(I_c^2)$. We compute this over all the neighboring communities. So the total time is $O(C_v^i) \cdot O(I_c^2)$.

However, the sum of $I_c$ from all the communities will be less than equal to $Rd_v$, where $R$ is the maximum number of times an edge is
shared between communities. Therefore, the time to update can be written as the square of sum of components that together add to $Rd_v$. 
Since sum of squares of elements is less than the square of the sum of the element, therefore, an upper bound on the time is $O({Rd_v}^2)$.

Therefore the total operation on $v$ is $O(TC_v^i) + O({Rd_v}^2)$, which can be rounded to the more expensive operation $O({Rd}^2)$. It is done over every vertex and for (say,) $Iter$ iterations.  So total time is $O(Iter \cdot n \cdot ({Rd}^2))$ (where $d$ is the average degree
of the vertices ans $n$ is the number of nodes).

In order to make a comparison, we also report the complexity of all the competing algorithms in Table ~\ref{com}.

\vspace{-3mm}

\begin{table}[!h]
\centering
\caption{ Time complexity of the competing overlapping community finding algorithms ($n$: number of nodes, $m$: number of edges, $T$: maximum number of iterations, $d$: average degree of the nodes).}\label{com}
\scalebox{0.90}{
 \begin{tabular}{|c|l|}
\hline
  Algorithm & Complexity \\\hline
OSLOM & $O(n^2)$ \\\hline
COPRA & $O(v^3n)$ plus $O(vn\ log (v))$ per iteration, where $v$ is\\ 
      &  maximum number of communities per vertex)\\\hline
SLPA  & $O(T\ nk)$ or $O(T\ m)$ \\\hline
EAGLE & $O(n^2 + (h + n)T)$, where $h$ is the number of cliques\\\hline
MOSES & $O(mn^2)$ \\\hline
BIGCLAM &  $O(m)$ per iteration \\\hline
MaxGenPerm & $O(n \cdot ({Rd}^2))$ per iteration, where $R$ is the \\
           &  maximum number of communities shared by an edge \\\hline 

 \end{tabular}
}
\end{table}

\vspace{-5mm}

 \begin{algorithm}[!htb]
\scriptsize
  \caption{MaxGenPerm: GenPerm Maximization}\label{algo}
        {{\bf Input:}  A connected graph $G = (V,E)$\\
        {\bf Output:}  Detected overlapping communities and GenPerm of $G$, $GenPerm\_G$ }
        
  \begin{algorithmic}
  \State \underline{{\bf Initialization}}:
  \State {Assign each edge (two end vertices) to a separate community}
  \State{$GenPerm\_G \gets 0.0$, $SumPerm \gets 0.0$}
  \State {$OldPerm \gets -1.0$}
  \State {Set the value of maximum number of allowable iteration as $MaxIter$}
  \State {$Iter \gets 0$}
  
  \State \underline{{\bf Update}}:
  \While { $Iter < MaxIter$ or $SumPerm \neq OldPerm$ }
  \State{$Iter \gets Iter + 1$}
  \State{$OldPerm \gets SumPerm$}
  
  \For {each vertex $v$}   
  
  \State {$CurComm$ is the set of communities to which $v$ belongs}
  \State{Find $P^{cur}_g(v)$, the GenPerm of $v$ in  $CurComm$} 
 
  \If {$P^{cur}_g(v)==1$}
  \State {$SumPerm \gets SumPerm+P^{cur}_g(v)$}
  \State{\Comment  $CurComm$ represents the highest GenPerm value of $v$.}
  \State {\Comment No need to continue further.}
  \Else

  \State {Determine $CNeigh$, the set of neighboring communities of $v$}
  \State{\Comment Find overlapping permanence of  $v$ in $CNeigh$}
  \State {$P^{temp}_g(v)\gets 0.0$}
 \State {$TempComm \gets \emptyset$}
  \For {each community $c$ in $CNeigh$}
  \State{Temporarily assign $v$ to community $c$}
 \State {Calculate $P_{g}^c(v)$, GenPerm of $v$ in $c$}
 
 \If {$P_{g}^c(v) > 0$}
 \State {$P^{temp}_g(v) \gets P^{temp}_g(v)+P_{g}^c(v)$}
 \State {$TempComm \gets TempComm \cup c$}
 \EndIf  
  \EndFor
  
   \State{\Comment  Update $CurComm$ if overall GenPerm is higher.}
 \If{$P^{temp}_g(v) > P^{cur}_g(v)$}
 \State {$P^{cur}_g(v) \gets P^{temp}_g(v)$}
 \State {$CurComm \gets TempComm$}
  \EndIf  
   \State {$SumPerm \gets SumPerm+P^{cur}_g(v)$}
    
      \EndIf 
  \EndFor

  \EndWhile 
  \State{$GenPerm\_G \gets SumPerm/|V|$} \Comment GenPerm  of the graph
   \State{{\bf return} $GenPerm\_G$}
  
 \end{algorithmic}
 \end{algorithm}

\begin{figure*}
\centering
\scalebox{0.45}{
\includegraphics{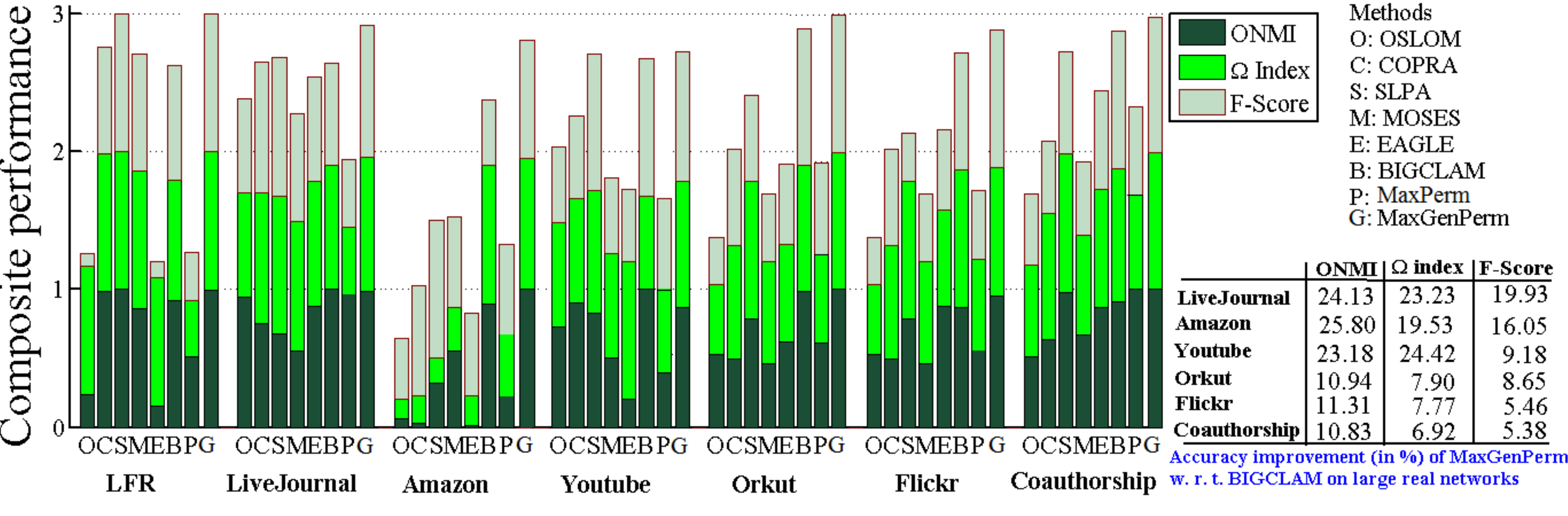}}
\caption{(Color online) Performance of various competing algorithms (ranging from 0 to 3) to detect the ground-truth communities.
The table shows the performance improvement of MaxGenPerm over BIGCLAM in detecting communities in large real
networks.}\label{composite_perf}
\end{figure*}

\vspace{-5mm}

\subsection{Performance Analysis}
In order to evaluate MaxGenPerm, we 
(i) compare the detected community with the ground-truth community and measure the similarity,  and 
(ii) check whether the algorithm is stable under different vertex orderings. 

\subsubsection {Comparison with ground-truth communities} 
\label{extrinsic}
We run MaxGenPerm along with six other algorithms mentioned in Section \ref{baseline} and compare their performance 
with networks whose ground-truth communities are known. We further consider MaxPerm (permanence maximization) \cite{chakraborty_kdd} as another competing method in order to see what extent it captures the underlying community structure. Since the baseline methods do not scale for
large-size real networks,  we use the sampled subnetworks as mentioned in Section \ref{sample}. For
the LFR benchmark, however, the results are reported on the entire networks.  In the interest of space, the
results are shown for the following setting of the LFR
network: $n$=1000, $\mu$=0.2, $O_n$=5\% and
$O_m$=4. For each real network, we measure the average value of each validation metric for 500 different samples.

For each validation metric (ONMI,  $\Omega$ Index, F-Score), we separately scale the scores of the methods so that
the best performing community detection method has the score of $1$. Finally, we compute the composite performance by summing up the
$3$ normalized scores. If a method outperforms all the other methods in all the scores, then its composite performance is $3$. 

Figure \ref{composite_perf} displays the composite performance of the methods for different networks (actual
values are reported in SI Text). On an average, the composite
performance
of MaxGenPerm (2.88) significantly outperforms other competing algorithms: 6.27\% higher than that of BIGCLAM (2.71), 18.03\% higher than that
of SLPA (2.44), 101.3\% higher than that of OSLOM (1.43), 36.4\% higher than that of COPRA (2.11), 48.4\% higher than that of MOSES (1.94), 70.41\% higher than that of MaxPerm (1.69), 
and 77.8\% higher than that of EAGLE (1.62). The absolute average ONMI of MaxGenPerm for one LFR and six real networks taken together is 0.85,
which is
4.93\% and 26.8\% higher than the two most competing algorithms, i.e., BIGCLAM\footnote{We did two experiments: first, we fixed the number of communities of BIGCLAM same as that in the ground-truth structure, and second, we retained the default values to the parameters used in BIGCLAM and it automatically computed the number of communities on the fly. We observed that the second case produced comparatively better results, and therefore we report these value here.} (0.81), and SLPA (0.67)
respectively. In terms of absolute values of
scores, MaxGenPerm achieves the average F-Score of 0.84 and average $\Omega$ Index of 0.83. Overall, MaxGenPerm gives the best results, 
followed by BIGCLAM, SLPA, COPRA, MOSES, EAGLE, MaxPerm and OSLOM.\\

\noindent {\bf Comparison with BIGCLAM for large networks:} As most of the baseline algorithms except BIGCLAM do not scale for large real
networks \cite{Leskovec}, we separately compare MaxGenPerm with 
BIGCLAM (which is also the most competing algorithm) on actual large real datasets. The table in Figure \ref{composite_perf} shows the
percentage
improvement of MaxGenPerm over BIGCLAM for different real networks. On average, MaxGenPerm achieves 17.67\% higher ONMI, 14.96\% higher
$\Omega$ Index, and 10.78\% higher F-Score. Overall, MaxGenPerm outperforms BIGCLAM in every measure and for every network. The absolute
values
of the scores of MaxGenPerm averaged over all
the networks are 0.81 (ONMI), 0.82 ($\Omega$ Index), and 0.81 (F-Score).  Therefore, the improvement of MaxGenPerm over BIGCLAM is higher
considering the entire network 
in comparison to that in the sampled networks. 

The running times\footnote{All the experiments that we report here were run on a
64bit Linux machine (Ubuntu 10.04.4 LTS) with 4X 2.27 GHz i3 CPU and 3GB of RAM.} (in seconds) of current implementation of MaxGenPerm (BIGCLAM\footnote{Note other baseline algorithms are not scalable and can not run on large networks.}) for large real networks are as follows: LiveJournal: 108,000 (86,198);  Amazon: 9,145 (7,632);  Orkut: 92,998 (72,806); Youtube: 31,650 (28,763); Flickr: 2,072 (1,876); and Coauthorship:
9,572 (7,983). One of the future task would be to make MaxGenPerm algorithm fast with efficient data structures.  \\

\begin{figure}
 \includegraphics[width=\columnwidth]{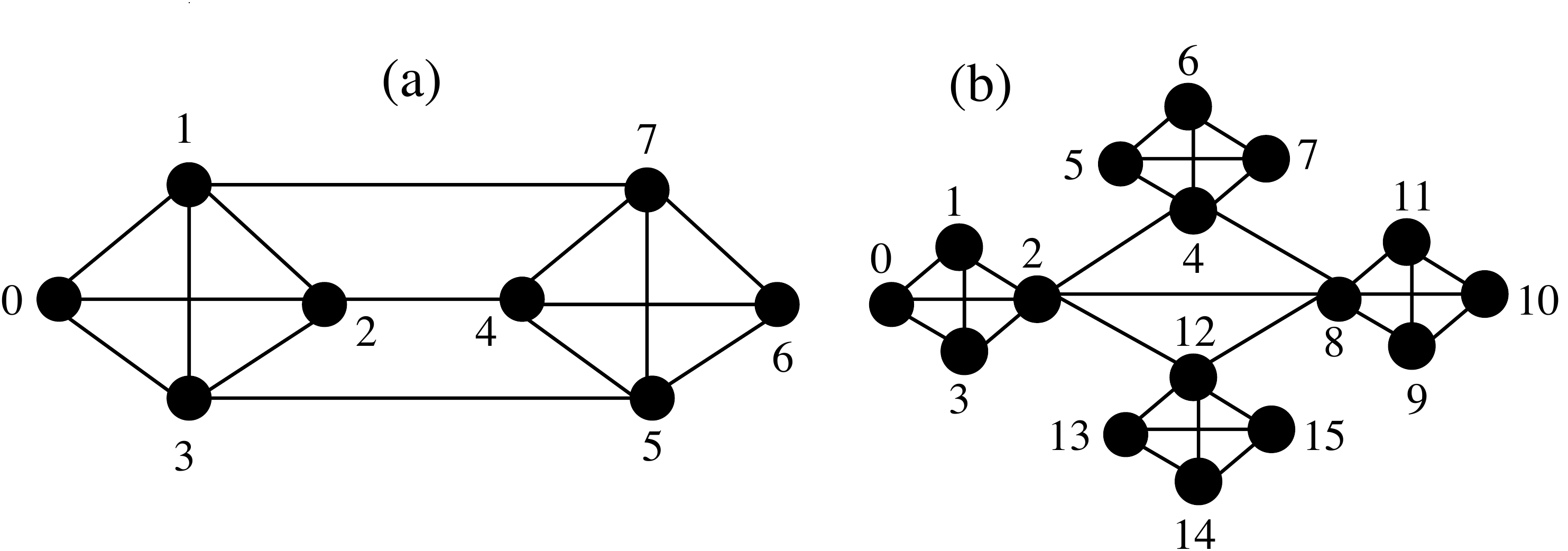}
\caption{Networks containing non-overlapping community structure.}\label{toy}
\end{figure}

\noindent{\bf Identifying non-overlapping communities:}
 As discussed earlier, all  real-world networks contain some degree of overlaps in their communities. Therefore, we generate two examples,
shown in Figure \ref{toy}, to demonstrate the ability of MaxGenPerm to identify non-overlapping groups. We compare MaxGenPerm with BIGCLAM
which was the closest competitor. For Figure \ref{toy}(a), BIGCLAM produces three community structures: \{0,1,2,3\}, \{4,5,6,7\},
\{1,2,3,5,4,7\}; whereas MaxGenPerm detects 
two
non-overlapping communities: \{0,1,2,3\}, \{4,5,6,7\}. Similarly, for Figure \ref{toy}(b), BIGCLAM detects six communities: \{0,1,2,3\},
\{4,5,6,7\}, \{8,9,10,11\}, \{12,13,14,15\}, \{2,4,8\}, \{2,8,12\}; whereas MaxGenPerm produces four disjoint communities: \{0,1,2,3\},
\{4,5,6,7\}, \{8,9,10,11\}, \{12,13,14,15\}. Note that one can obtain the same community structure by using MaxPerm (permanence maximization) \cite{chakraborty_kdd} as obtained from MaxGenPerm. Although these examples are on idealized networks, this experiment demonstrates that MaxGenPerm  can detect
non-overlapping communities, without  including non-intuitive overlapping structures.

\subsubsection{Effect of vertex ordering } \label{ordering}
Most of the community detection algorithms are based on optimizing certain functions 
(such as modularity), and the values are heavily dependent on the order in which vertices are processed \cite{chakraborty,lf2012}. 
Hence an intrinsic goodness of an algorithm can be measured in terms of its (resistant to) change in 
output (communities)  with change in initial vertex ordering. Such changes in community structure due to vertex ordering have been observed
earlier in the case of non-overlapping communities \cite{chakraborty,lf2012}. 
This can be more specifically measured in terms of number of invariant groups of vertices which stay in the same output 
community in spite of the fluctuation. These groups have been termed as ``constant communities'' \cite{chakraborty}.
In our earlier work, we showed that despite such fluctuations in the final outcome, there exist few such constant communities which always remain same across different vertex orderings. 

We measure the ratio ($\phi$) of
the number of constant communities to the total number of nodes.
 In the worst case, the number of constant communities would be equal to the number of nodes with each node being a community. 
 If the value of $\phi$ for an algorithm
remains constant and small  over different vertex orderings, then the algorithm is less sensitive to vertex ordering. 

\begin{figure}
\centering
  \includegraphics[width=\columnwidth]{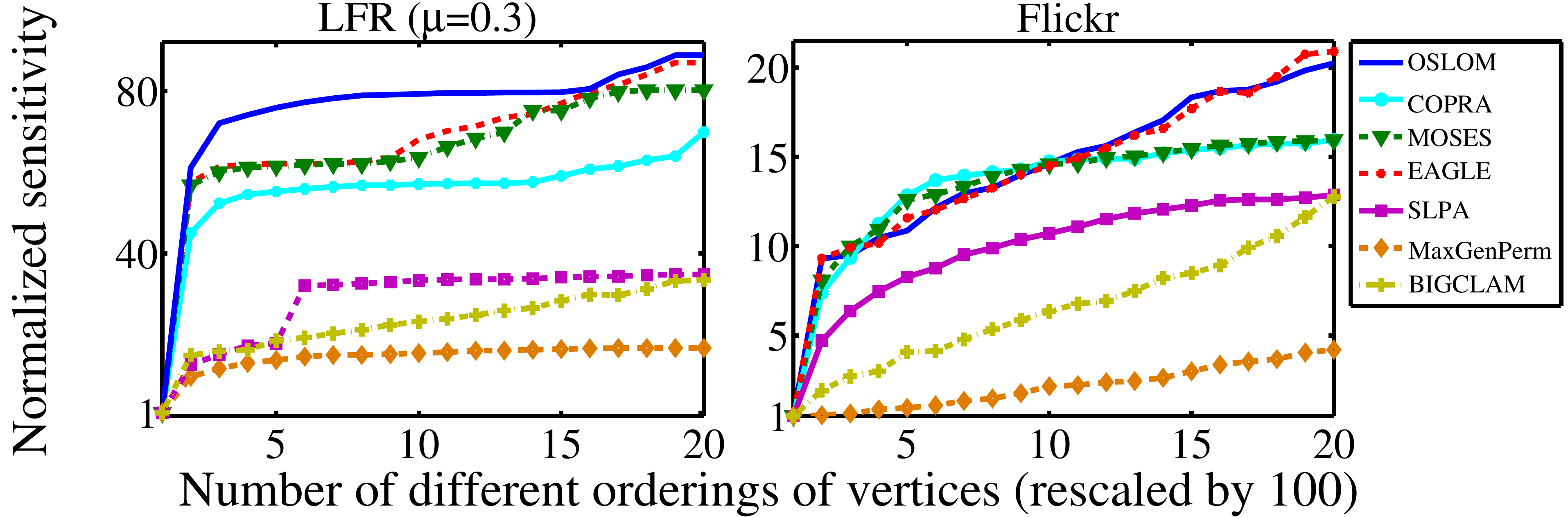}
 \caption{(color online) The $\phi$ value for each algorithm across 2000 different vertex orderings. The x-axis is rescaled by a constant
factor of
100. For better visualization, we rescale $\phi$ with the
minimum
value for each algorithm so that the sensitivity of all the algorithms starts from 1.}\label{constant_comm}
\end{figure}

By comparing the value of
$\phi$ for different vertex orderings (Figure \ref{constant_comm}), we observe that MaxGenPerm is the least sensitive algorithm, followed
by BIGCLAM, SLPA, COPRA, MOSES, EAGLE and OSLOM.  These results unfold another major characteristic of our algorithm that by producing lesser number of competing solutions, our algorithm is able to significantly reduce the problem of
``degeneracy of solutions'' \cite{Fortunato}.

\begin{figure}[!h]
 \centering
 \scalebox{0.25}{
 \includegraphics{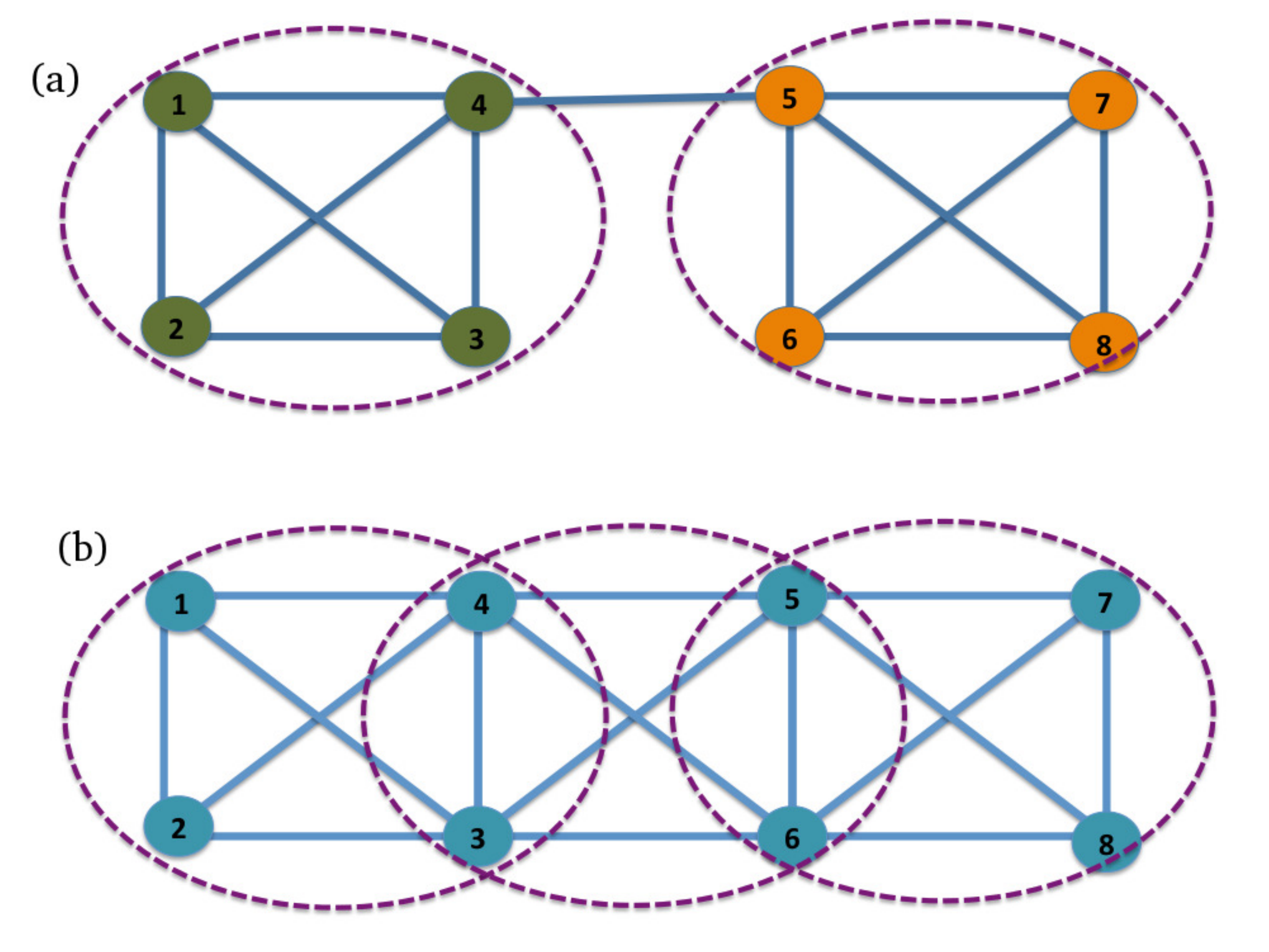}}
 \caption{(Color online) Examples comparing the behavior of permanence and GenPerm. Vertices of the same color represent the communities given by maximizing permanence. Vertices within the same dotted circle represent the communities given by maximizing GenPerm.} \label{3clq}
 \end{figure}

\section{Mitigation of Resolution Limit by GenPerm}\label{discussion}
Resolution limit  \cite{Fortunato,good2010} occurs when communities smaller than a certain size get merged into larger communities. In ~\cite{chakraborty_kdd}, we theoretically proved that communities obtained by maximizing permanence can mitigate the resolution limit problem. However, maximizing permanence still experiences resolution limit  if a vertex is tightly connected to more than one community. In such cases, finding communities by maximizing GenPerm provides a better solution.

Figure \ref{3clq}(a) shows two loosely connected communities. In this case, both permanence
and GenPerm give the same communities upon maximization. Figure \ref{3clq}(b) shows tightly connected communities. Maximum permanence is
obtained
when all the vertices are in the same community and therefore fails to overcome the resolution limit. In contrast, maximizing GenPerm,
produces three overlapping communities. \\

\noindent{\bf Resolution limit in overlapping communities:}  The example discussed above demonstrates that maximizing GenPerm can mitigate
the resolution limit for both
non-overlapping and overlapping communities. To the best of our knowledge, resolution limit in overlapping communities has not been studied 
before.  We define the following two criteria for identifying small but distinct groups of vertices in overlapping communities.
\begin{enumerate}
\item Small communities that are loosely connected to larger communities will  remain distinct and do not fall into the overlap between the
larger communities
\item If all the vertices in a community belong to overlap regions, the community should still be identified as a separate community by itself. 
\end{enumerate}

\begin{figure*}
\centering
\scalebox{1}{
\includegraphics[width=0.9\textwidth]{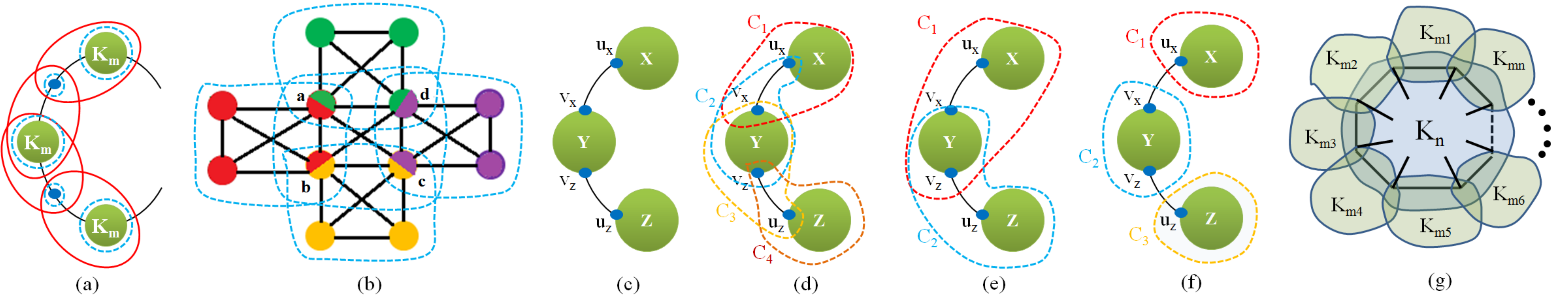}}
\caption{(Color online) Toy examples used to demonstrate the theorems ($K_x$: clique of size $x$).}
\label{theorem}
\end{figure*}

These criteria are illustrated in Figure \ref{theorem}.
 Figure \ref{theorem}(a) shows a circle of cliques, where two consecutive cliques are connected by a bridging
vertex.
All
the baseline algorithms  
output communities where the bridging vertex is an overlapping vertex and
part of two
neighboring cliques (shown by the community structure marked through solid lines). However, maximum value of GenPerm is obtained when each
bridging vertex forms a separate singleton community along with the individual cliques (shown by the community structure marked through
broken lines), as required by Criterion 1.

In Figure \ref{theorem}(b), we have a central clique, and four other cliques sharing one edge each with the central clique.
Most baseline algorithms detect four communities (indicated by the four colors) and vertices $a$, $b$, $c$ and $d$ are treated as overlapping
vertices. However, there are actually five cliques including the one in the center (shown by the community structure marked by
the broken lines). Maximizing GenPerm is able to detect
these five cliques as separate communities, with overlap at the corners of the central clique. This satisfies the second criterion.

We can generalize these  examples  to the following two
theorems, over  cliques of  any given size as follows (see SI Text for detailed discussion):

THEOREM 1. {\it Given three cliques $X$, $Y$ and $Z$, such that $X$ and $Z$ are not connected to each other and $Y$ is connected to $X$
and $Z$ by two
edges ($u_x, v_x$)  and ($u_z, v_z$) respectively where $u_x \in X$, $u_z \in Z$ and $v_x, v_z \in Y$, the highest GenPerm
 is
obtained, if $X$, $Y$ and $Z$ are three separate communities with no overlap.}

\begin{proof}

We prove the theorem using the example given in Figure~\ref{theorem}(c). The detailed proof is mentioned in SI Text. Let the size of cliques $X$, $Y$ and $Z$ be $n_x$, $n_y$ and $n_z$ respectively. We consider the following scenarios where $Y$ forms an overlap with  $X$ and $Z$. In every scenario, we only consider the sum of GenPerm of  $u_x$, $v_x$, $u_z$ and $v_z$ because all other vertices are unaffected by any kind of
community assignment and have GenPerm equal to 1.  

In Case 1 (shown in Figure~\ref{theorem}(d)), $X$ has an overlap with part of $Y$ (community $C_1$), $Y$ has an overlap with part of  $X$
(community $C_2$), $Y$ also has an overlap with part of $Z$ (community $C_3$) and $Z$ has an overlap with part of $Y$ (community $C_4$). The
overlaps are such that $u_x \in C_1$, $ u_x \in C_2$; $v_x \in C_1$, $v_x \in C_2$, $ v_x \in C_3$; $v_z \in C_2$, $v_z \in C_3$, $v_z
\in C_4$; $u_z \in C_3$, $ u_z \in C_4$. This represents the most general case of overlap between the cliques.

Note that all the vertices in the network except $u_x$, $v_x$, $u_z$ and $v_z$ are unaffected for any kind of
community assignment and they have $P_{g}$ as 1.

Vertex $u_x$ has no external connection in $C_1$ and $C_2$. In that case, we assume $E_{max}=1$ to avoid the ``divide by zero'' case  for
computing $P_{g}$. Therefore, $P_{g}(u_x)=P_{g}^{C_1}(u_x)+P_{g}^{C_2}(u_x)=\frac{(n_x-2)(2n_x-1)}{2n_x^{2}}$.

Similarly for $v_x$, there is no external connection in any of its assigned communities $C_1$, $C_2$ and $C_3$. Therefore, $P_{g}(v_x)=\frac{2n_y-3}{2n_y}$.
The connectivity of $v_z$ is similar to $v_x$. Therefore, $P_{g}(v_z)=\frac{2n_y-3}{2n_y}$. Further, the connectivity of vertex $u_z$ is similar to that of $u_x$ except the degree of $u_z$ being $n_z$. Therefore, $P_{g}(u_z)=\frac{(n_z-2)(2n_z-1)}{2n_z^{2}}$.

Now combining GenPerm of the affected vertices, we obtain:
\begin{equation}\label{case1}
\begin{split}
 P_{g}^1=P_{g}(u_x)+P_{g}(v_x)+P_{g}(v_z)+P_{g}(u_z)\\
=4-\frac{5}{2n_x}-\frac{5}{2n_z}-\frac{3}{n_y}+\frac{1}{n_x^2}+\frac{1}{n_z^2}
\end{split}
\end{equation}

In Case 2, as shown in Figure~\ref{theorem}(e), $X$ has complete overlap with $Y$, and $Z$ also has a complete overlap with $Y$. Therefore $u_x
\in C_1$; $v_x \in C_1$, $v_x \in C_2$; $v_z \in C_1$, $v_z \in C_2$; $u_z \in C_2$.

For vertex $u_x$ which belongs to community $C_1$ only, there is no external connection, and no internal edges are shared. Therefore, $P_{g}(u_x)=P_{g}^{C_1}(u_x)=\frac{n_x-2}{n_x}$. For vertex $v_x$ belonging to both $C_1$ and $C_2$, the internal edges in clique $Y$ are shared by two communities. So,  $P_{g}(v_x)=\frac{n_y^2-n_y-1}{n_y^2}$. Similarly, GenPerm of $v_z$ is similar to that of $v_x$. Therefore, $P_{g}(v_z)=\frac{n_y^2-n_y-1}{n_y^2}$.

The connectivity of $u_z$ is similar to that of $u_x$, except the degree of $u_z$ being $n_z$. Therefore,  $P_{g}(u_z)=\frac{n_z-2}{n_z}$.

Now combining GenPerm of all the affected vertices, we obtain:
\begin{equation}\label{case2}
\begin{split}
 P_{g}^2=P_{g}(u_x)+P_{g}(v_x)+P_{g}(v_z)+P_{g}(u_z)\\
 =4-\frac{2}{n_x}-\frac{2}{n_y}-\frac{2}{n_z}-\frac{2}{n_y^2}
\end{split}
\end{equation}

In Case 3, as shown in Figure \ref{theorem}(f), each of $u_x$,
$v_x$, $v_z$ and $u_z$ have one external neighbor and the internal clustering coefficient is 1. Therefore,
$P_{g}(u_x)=P_{g}^{C_1}(u_x)=\frac{n_x-1}{n_x}-(1-1)\cdot\frac{n_x-1}{n_x}=\frac{n_x-1}{n_x}$.
Similarly, $P_{g}(v_x)= \frac{n_y-1}{n_y}$, $P_{g}(v_z)=\frac{n_y-1}{n_y}$ and $P_{g}(u_z)=\frac{n_z-1}{n_z}$.

Therefore, combining GenPerm of the affected vertices, we obtain:
\begin{equation}\label{case3}
\begin{split}
 P_{g}^3=P_{g}(u_x)+P_{g}(v_x)+P_{g}(v_z)+P_{g}(u_z)\\
=4-\frac{1}{n_x}-\frac{2}{n_y}-\frac{1}{n_z}
\end{split}
\end{equation}

When comparing $P_{g}^1$, $P_{g}^2$ and $P_{g}^3$, the last scenario gives the highest GenPerm value. Thus the theorem is
proved.
\end{proof}

THEOREM 2. {\it Given a central clique $X$ and $n$ surrounding cliques such that each shares exactly one edge with $X$, the
highest
GenPerm is obtained if each of the surrounding cliques and $X$ are identified as a separate community (total $n+1$
communities) with the vertices of $X$ forming an overlap with $X$ and the two other communities to which it is connected.}

\begin{proof}
Let us assume the toy example shown in Figure \ref{theorem}(b), a generalized version of which is shown in Figure \ref{theorem}(g). Assume
that there is a central clique $X$ of size $n$ which is surrounded
by $n$ cliques, $C_1$, $C_2$, ...., $C_n$ with size $n_1$, $n_2$,..., $n_n$ respectively such that each of the surrounding cliques share one
edge with
$X$.

Now let us assume the community structure as shown by the broken lines in Figure \ref{theorem}(b) where there are $n+1$
communities, namely $X$, $C_1$, $C_2$, ...., $C_n$, and the nodes in $X$ such as $a$, $b$, $c$, $d$, $...$ are overlapping nodes and each
of them belong to two 
communities (we call these overlapping nodes together as set $V_{ov}$). One can measure GenPerm of the
graph in this setting. It is simple to show that the vertices except $V_{ov}$ have $P_{g}$ as 1.  However, we calculate $P_{g}$ of each
vertex in $V_{ov}$ as follows. Note that each node in $V_{ov}$ has equal $P_{g}$ value.

Let us consider vertex $a$ ($\in V_{ov}$) which shares $C_1$, $C_2$ and $X$ communities. The degree of $a$ is $D(a)=n_1+n_2+n-5$. In
community $C_1$, $a$ has $n_1-2$ internal connections which
are not shared, but edge ($a,d$) is shared with both $C_1$ and $X$. It has no external connection and its internal clustering coefficient in
this community is 1. Therefore, $P_{g}^{C_1}(a)=\frac{n_1-3/2}{n_1+n_2+n-5}$. Similarly, $P_{g}^{C_2}(a)=\frac{n_2-3/2}{n_1+n_2+n-5}$.

In community $X$, $a$ has $n-1$ internal connections among which two edges, namely ($a,b$) and ($a,d$) are part of two communities.
It has no external connection and the internal clustering coefficient is 1. Therefore, $P_{g}^X(a)=\frac{n-2}{n_1+n_2+n-5}$.

Combining all these cases, we obtain the overall GenPerm of vertex $a$ as $P_{g}(a)= \frac{n_1+n_2+n-5}{n_1+n_2+n-5}=1$.

Similarly, one can show that GenPerm of each vertex in $V_{ov}$ is 1. Since the maximum value of $P_{g}$ for a vertex
is 1, this configuration is the best community assignment one can obtain using maximizing $P_{g}$. Therefore,
this arrangement gives the highest GenPerm.
\end{proof}

\vspace{-5mm}

\section{Related Work}\label{related_work}
There exist many different methods for identifying non-overlapping communities. For details on different community detection methods,
readers are referred to following reviews: \cite{porter09}.

When the true community structure is not known, a quality function such as modularity \cite{NewGir04} is used to evaluate the performance of
the clustering algorithms. Shen et al. \cite{Shen} introduced a variant of modularity for overlapping communities, which was
later modified by L{\'a}z{\'a}r  et al. \cite{Vicsek}.  Shen et al. \cite{Hua-We} proposed another variant on the
basis that a maximal clique only belongs to one community. Nicosia et al. \cite{Nicosia} extended the definition to directed
graphs with overlapping communities. Chen et al. \cite{Chen} proposed an extension of modularity density for overlapping community
structure. Several other extensions of modularity \cite{ chen2010detecting, Gregory} have also  been proposed. It is worth mentioning that the formulation of GenPerm  is an extension of our earlier work, where we proposed {\em permanence}, a new scoring metric for non-overlapping community \cite{chakraborty_kdd}.

There has been a class of algorithms for network clustering, which allow nodes
belonging to more than one community. Palla proposed
``CFinder'' \cite{PalEtAl05}, the seminal and most popular method based on clique-percolation technique. However, due to the clique
requirement and the sparseness of real networks, the communities discovered by CFinder are usually of low quality \cite{Fortunato:2009}. The
idea of partitioning links instead of nodes to discover community structure has also been explored \cite{nature2010}

On the other hand, a set of algorithms utilized local expansion and optimization to detect overlapping communities. For instance, Baumes
et al. \cite{BaumesGKMP05} proposed ``RankRemoval'' using a local density function. 
MONC \cite{abs-1012-1269} uses
the modified fitness function of LFM which allows a single node to be considered a community by itself. OSLOM \cite{oslom} tests the
statistical significance of a cluster with respect to a global null model (i.e., the random graph generated by the configuration model)
during community expansion. Chen et al. \cite{chen2010detecting} proposed selecting a node with maximal node strength based on two
quantities -- belonging degree and the modified modularity. EAGLE \cite{Shen} and GCE \cite{lee} use the agglomerative framework to produce
overlapping communities.

Few fuzzy community detection algorithms have been proposed that quantify the strength of association between all pairs of
nodes and communities \cite{Gregory}. Nepusz et al. \cite{Nepusz} modeled the overlapping community detection as a nonlinear constrained
optimization problem which can be solved by simulated annealing methods. Due to the probabilistic nature, mixture models provide an
appropriate framework for overlapping community
detection \cite{Newman05062007}. 
 MOSES \cite{moses} uses a local optimization scheme in
which
the fitness function is defined based on the observed condition distribution. 
Ding
et al. \cite{ding} employed the affinity propagation clustering algorithm for overlapping community detection. Whang et al. \cite{Whang:2013} developed an overlapping community detection algorithm using a seed set expansion approach.  Recently, BIGCLAM \cite{Leskovec}
algorithm is also built on NMF framework.  

The label propagation algorithm has been extended to overlapping community detection by allowing a node to have multiple labels. In COPRA
\cite{Gregory1}, each node updates its belonging coefficients by averaging the coefficients from all its neighbors at each time step in a
synchronous fashion. SLPA \cite{Xie, abs-1105-3264} spreads labels between nodes according to pairwise interaction rules. A game-theoretic
framework is proposed by Chen et al. \cite{Chen:2010} in which a community is associated with a Nash local equilibrium.

Beside these, 
Zhang et al. \cite{ZhangWWZ09} proposed an iterative process that reinforces the network topology and proximity that is interpreted as the
probability of a pair of nodes belonging to the same community. Istv{\'a}n et al. \cite{pone.0012528} proposed an approach focusing on
centrality-based influence functions. Gopalan  and Blei \cite{Gopalan03092013} proposed an algorithm that naturally interleaves
subsampling from the network and updating an estimate of its communities. 
Recently, Sun et al. \cite{Sun20111060}
proposed fuzzy clustering based non-overlapping community detection technique that can be extended to overlapping case.
However, none of these algorithms can work equally well for both overlapping and non-overlapping cases.

\section{Conclusion and Future Work}\label{end}

We proposed a novel vertex-based metric, called GenPerm which provides a quantitative measure of how much a vertex belongs
to each of its constituent communities, and to the whole network. Experimental results showed that our metric qualifies as a better
community scoring metric. Our work has several implications: First, our metric is a generalized formulation that can be used evaluate the
quality of a community structure. Second, MaxGenPerm is the first generalized algorithm that can identify both overlapping and
non-overlapping communities, without any prior information. Third, to the best of our knowledge, the issue related to resolution limit in
the context of overlapping community structure is discussed for the first time in this work. We showed that maximizing GenPerm, quite
significantly, reduces the effect of resolution limit. Fourth, we showed that GenPerm suitably explains the position of vertices within
a community, and provides a ranking scheme of vertices that has been shown to be effective for initiator selection in message spreading.
Finally, being a vertex-based metric, overlapping permanence is fine-grained,
therefore allows partial estimation of communities in a network whose entire structure is not known.

We plan to extend our metric to evaluating dynamic communities and for detecting local communities around a specified vertex. We also plan
to explore other uses of GenPerm such as in identifying critical vertices during spread of epidemics.

%

\begin{IEEEbiography}[{\includegraphics[width=1in,height=1.25in,clip,keepaspectratio]{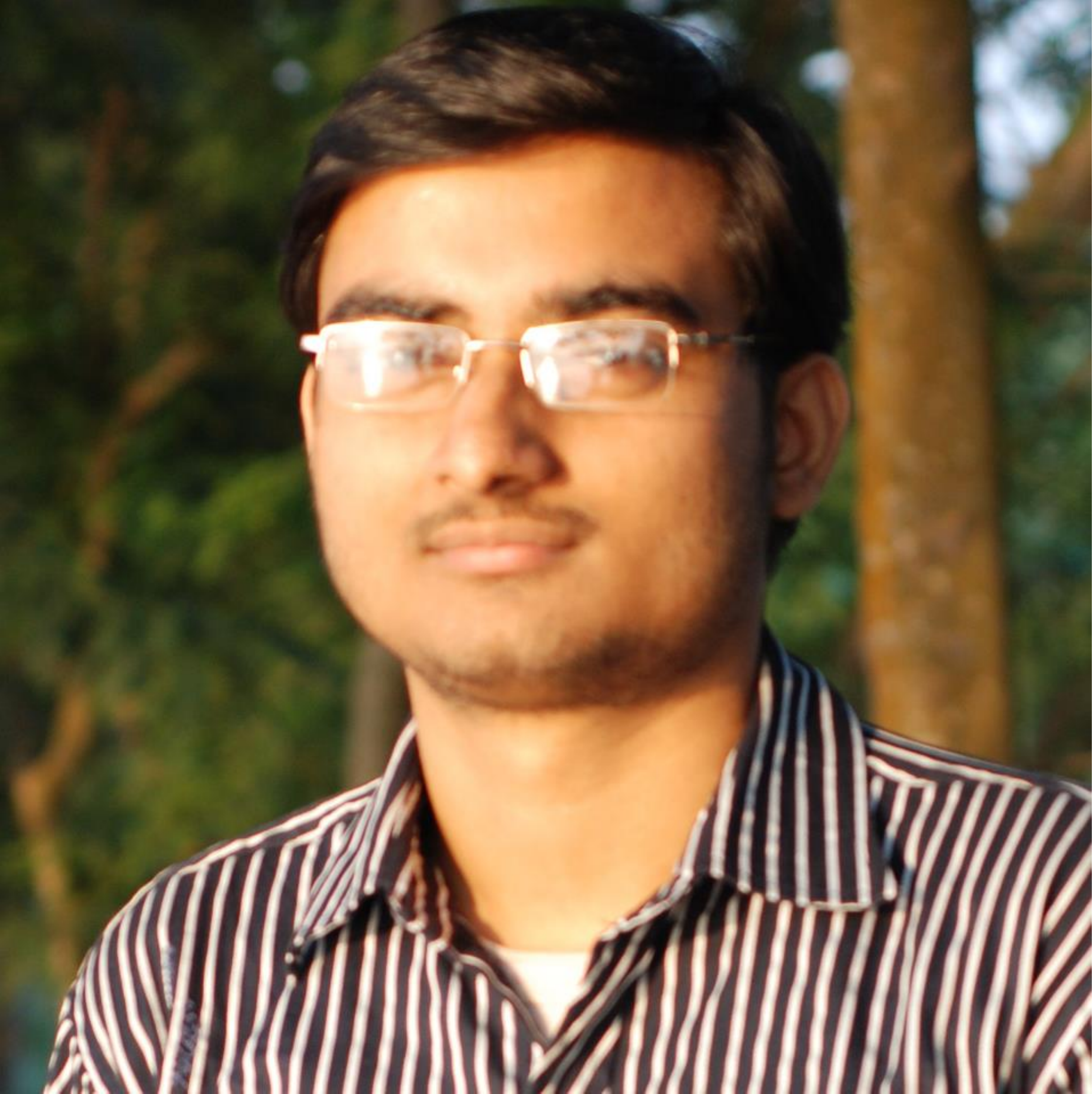}}]{Tanmoy Chakraborty}
is currently a postdoctoral fellow at Dept. of Computer Science, University of Maryland, College Park, USA. Prior to this, he completed his Ph.D. as a  Google India fellow from Dept. of CSE, IIT Kharagpur, India in 2015. His primary research interests include
Complex Networks, Social Media, Data Mining and NLP. More details: \url{https://sites.google.com/site/tanmoychakra88/}
\end{IEEEbiography}

\begin{IEEEbiography}[{\includegraphics[width=1in,height=1.25in,clip,keepaspectratio]{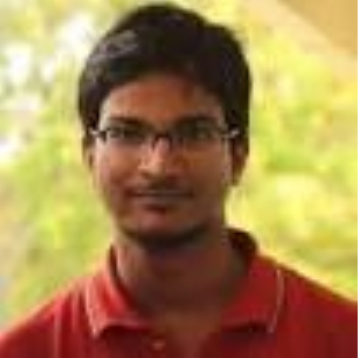}}]{Suhansanu Kumar}
is a graduate student in the Department of Computer Science, University of Illinois at Urbana-Champaign. He received his B.Tech degree in 2014  from IIT Kharagpur, India. His primary research interests include Social Networking, Machine Learning and Game Theory. More details: \url{http://illinois.academia.edu/SuhansanuKumar}.
\end{IEEEbiography}

\begin{IEEEbiography}[{\includegraphics[width=1in,height=1.25in,clip,keepaspectratio]{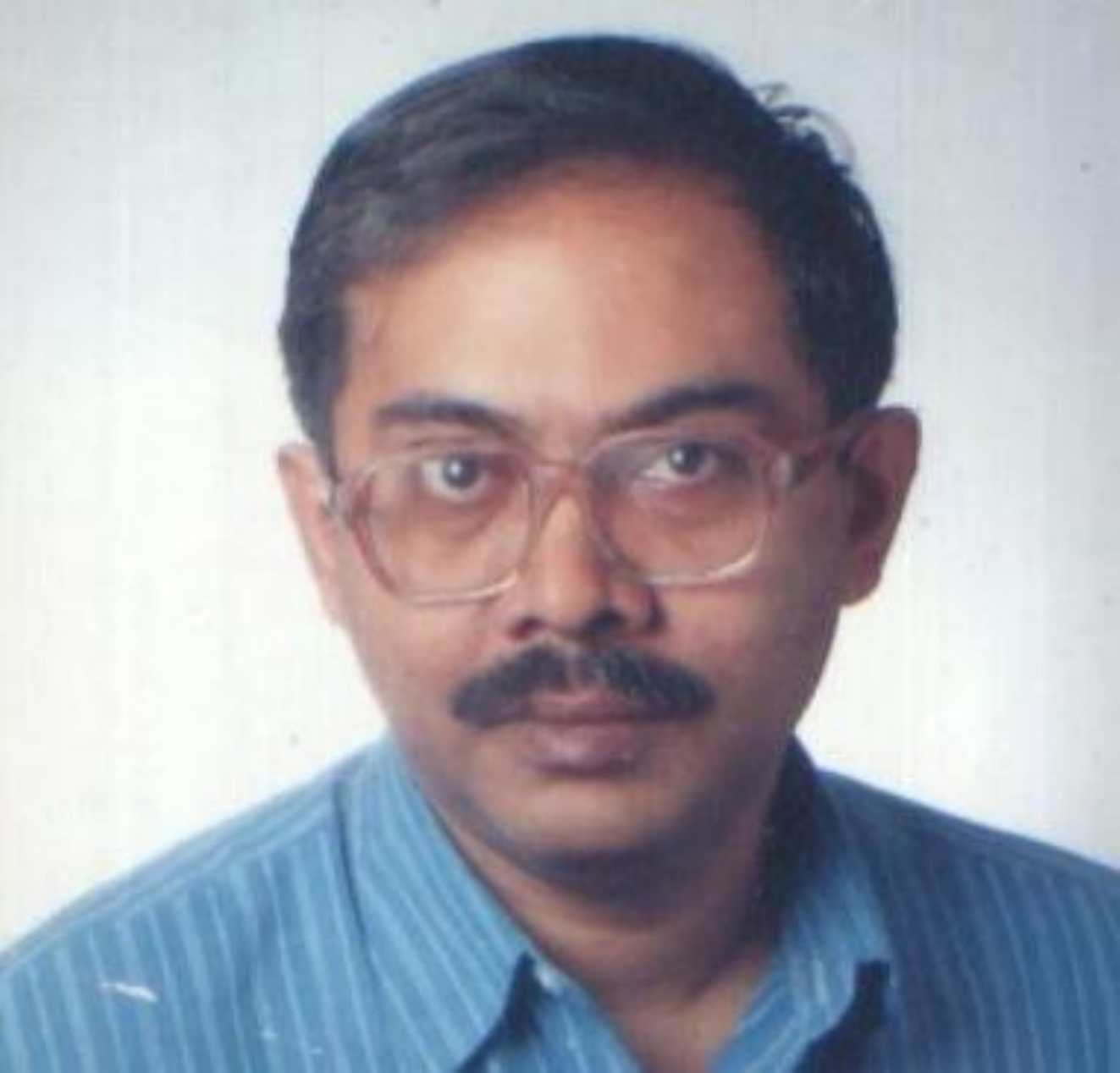}}]{Niloy Ganguly}
is a Professor in the Department of CSE, IIT Kharagpur, India. His primary research interests include Online Social Networking, Peer to peer
networking and Machine Learning.  More details: \url{http://www.facweb.iitkgp.ernet.in/~niloy/}.
\end{IEEEbiography}

\begin{IEEEbiography}[{\includegraphics[width=1in,height=1.25in,clip,keepaspectratio]{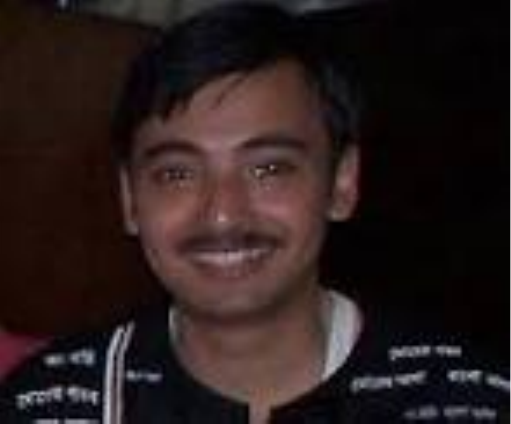}}]{Animesh Mukherjee}
 is an Assistant Professor in the Department of CSE, IIT Kharagpur, India, and a Simons Associate in the Abdus Salam International Centre
for Theoretical Physics, Trieste, Italy. His main research interests center around applying complex system approaches to different problems
in (a) human language evolution,  web social media,  and NLP.  More details: \url{cse.iitkgp.ac.in/~animeshm/}.
\end{IEEEbiography}

\begin{IEEEbiography}[{\includegraphics[width=1in,height=1.25in,clip,keepaspectratio]{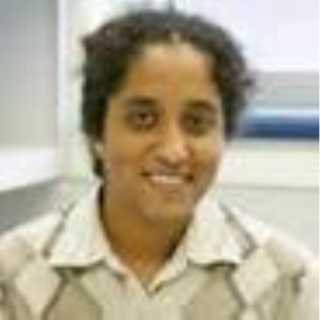}}]{Sanjukta Bhowmick}
is an Associate Professor in the Department of Computer Science, University of Nebraska, Omaha. Her research interests include Network
Analysis, HPC and Graph Theory. More details: \url{http://faculty.ist.unomaha.edu/sbhowmick/}.
\end{IEEEbiography}


\newpage

 {\Large\textbf{Supplementary Materials}}

\section{Sampled networks}
Since all the baseline algorithms (except BIGCLAM) are not suitable to run on the large networks, we adopt a community-centric sampling strategy \cite{Leskovec} to get small networks from each real-world network. The strategy is described in Section 3.1 of the main text. Here we report the properties of the sampled networks corresponding to each real-world network in Table \ref{pro}. 

\begin{table}[!h]
\centering
\caption{Average properties of the sampled real-world networks. $\bar N$: avg. number of nodes, $\bar E$: avg. number of edges, $\bar C$:
avg. number  of communities, $\bar \rho$: avg. edge-density per community, $\bar S$: avg. size of a community, $\bar O_m$: avg.
number of community memberships per node.}\label{pro}
\scalebox{0.90}{
 \begin{tabular}{c|r|r|r|r|r|r}
\hline
  Network &  $\bar N$ & $\bar  E$ & $\bar C$ &  $\bar \rho$ & $\bar S$ & $\bar O_m$ \\\hline
LiveJournal & 874.50 & 2463.09 & 96.73 & 0.32 & 10.34 & 4.80  \\
Amazon & 357.93 & 1345.45 & 12.76 & 0.34 & 14.34 & 15.56   \\
Youtube & 856.78 & 3180.34 & 261.94 & 0.12 & 13.63 & 5.19 \\
Orkut  & 623.43 & 2645.12 & 87.54 & 0.45 & 7.54 & 48.34 \\
Flickr & 763.21 & 2897.65 & 88.74 & 0.24 & 8.98 & 8.76 \\
Coauthorship & 976.76 & 3219.87 & 145.87 & 0.19 & 12.55 & 9.45    \\\hline
 \end{tabular}
}
\end{table}

\section{Experimental Setup}

\subsection{Community validation metrics}
The availability of ground-truth communities allows us to quantitatively evaluate the performance of community detection algorithms. For
evaluation, we use metrics that quantify the level of correspondence between the detected and the ground-truth communities. Given a network
$G(V, E)$, we consider a set of ground-truth communities $C$ and a set of detected communities $C^*$ where each ground-truth community $C_i
\in C$ and each detected community $C_i^* \in C^*$ is defined by a set of its member nodes. To quantify the level of correspondence of $C$
to $C^*$, we consider the following three validation metrics.

\begin{itemize}
 \item {\bf Overlapping Normalized Mutual Information (ONMI)} \cite{McDaid} adopts the criterion used in information theory to compare the
detected communities and the ground-truth communities. We use the ONMI implementation written by the authors and is available at
\url{https://github.com/aaronmcdaid/Overlapping-NMI}.

\item {\bf Omega ($\Omega$) Index} \cite{Gregory} is the accuracy on estimating the number of communities that each pair of nodes shares:
\begin{equation}
 \frac{1}{|V|^2} \sum_{u,v\in V} 1 \{|C_{uv}| = |C^*_{uv}|\}
\end{equation}

where $C_{uv}$ is the set of ground-truth communities that $u$ and $v$ share, and $C^*_{uv}$ is the set of detected communities that 
$u$ and $v$ share.

\item {\bf F-Score} measures a correspondence between each detected community to one of the ground-truth communities. To compute it, we
need to determine which $C^*_i \in C^*$ corresponds to which $C_i \in C$. We define F-Score to be the average of the F-Score of the
best-matching ground-truth community to each detected community, and the F-score of the best-matching detected community to
each ground-truth community as follows:
\begin{equation}
 \frac{1}{2}(\frac{1}{|C|} \sum_{C_i \in C} F(C_i,C^*_{g(i)})   +   \frac{1}{|C^*|} \sum_{C^*_i \in C^*} F(C_{g^{'}(i)},C^*_i)   )
\end{equation}
where the best matching $g$ and $g^{'}$ is defined as follows: $g(i)=\operatorname*{arg\,max}_j F(C_i, C^{*}_{j})$, 
$g^{'}(i)=\operatorname*{arg\,max}_j F(C_j, C^{*}_{i})$, and $F(C_i,C^*_j)$ is the harmonic mean of Precision and
Recall.

\end{itemize}

\begin{table*}[!ht]
\centering

\caption{Performance of various overlapping community detection algorithms in terms of three validation measures to detect ground-truth
communities from six real-world networks.}\label{result_real}
\scalebox{0.9}{
 \begin{tabular}{|c||c|c|c||c|c|c||c|c|c|}
 \hline
\multirow{2}{*}{\bf Algorithms} & \multicolumn{3}{c||}{{\bf LiveJournal}} & \multicolumn{3}{c||}{{\bf Amazon}} & \multicolumn{3}{c|}{{\bf
Orkut}} \\\cline{2-10}
  & {\bf ONMI} & {\bf $\Omega$ Index} & {\bf F-Score}   & {\bf ONMI} & {\bf $\Omega$ Index} & {\bf F-Score}  & {\bf ONMI} & {\bf
$\Omega$ Index} & {\bf F-Score} \\\hline

OSLOM & 0.741 & 0.609 & 0.557 & 0.052 & 	0.131 & 	0.388 & 0.643 & 0.671 & 0.528 \\\hline
COPRA & 0.593 & 0.760 & 0.775 & 0.028 &	0.190 & 	0.689 & 0.757 & 0.676 & 0.569 \\\hline
SLPA  & 0.535 & {\bf 0.801} & {\bf 0.821} &	0.273 & 0.173 &	0.865 & 0.731 & 0.788 & 0.947 \\\hline
MOSES & 0.433 & 0.756 & 0.641 &	0.473 &	0.301 &	0.566 & 0.447 & 0.667 & 0.528 \\\hline
EAGLE & 0.690 & 0.728 & 0.622 & 0.010 & 	0.209 &	0.518 & 0.177 & 0.487 & 	0.502 \\\hline
BIGCLAM & 0.780 & 0.719 & 0.606 & 0.766 & {\bf 0.954} & 0.412 & 0.738 & 0.598 & {\bf 0.958} \\\hline
MaxGenPerm & {\bf 0.789} & 0.775 & 0.782 & {\bf 0.854} &  0.904 & {\bf 0.887} & {\bf 0.767} & {\bf 0.812}  & 0.904 \\\hline
\multicolumn{7}{c}{}\\
\hline
\multirow{2}{*}{\bf Algorithms} & \multicolumn{3}{c||}{{\bf Youtube}} & \multicolumn{3}{c||}{{\bf Flickr}} & \multicolumn{3}{c|}{{\bf
Coauthorship}} \\\cline{2-10}
& {\bf ONMI} & {\bf $\Omega$ Index} & {\bf F-Score}   & {\bf ONMI} & {\bf $\Omega$ Index} & {\bf F-Score}  & {\bf ONMI} & {\bf
$\Omega$ Index} & {\bf F-Score} \\\hline
OSLOM & 	0.436 & 	0.426 & 	0.307 & 0.489 & 0.464 & 0.279 & 0.416 & 	0.521 & 	0.403 \\\hline  
COPRA & 0.410 & 	0.683 & 	0.639 & 0.459 & 0.745 & 0.583 & 0.521 & 0.712 & 	0.409 \\\hline
SLPA & 0.650  & 	{\bf 0.832} & 0.570 & 0.729 & {\bf 0.908} & 0.292 & 0.799 & 0.772 & 0.580 \\\hline
MOSES & 0.387 & 	0.612 & 	0.448 & 0.433 & 0.668 & 0.409 & 	0.548  & 0.562 & 0.418\\\hline
EAGLE & 0.516 & 	0.584 & 	0.531 & 0.816 & 	0.637 &	0.484 & 	0.711 & 	0.665 & 	0.556 \\\hline
BIGCLAM & 0.821 & 0.759 & 0.801 & 0.806 & {\bf 0.908} & 0.703  & 0.740 & 0.756 & 0.768\\\hline
MaxGenPerm	 & {\bf 0.831} & 0.821 & {\bf 0.912} & {\bf 0.883} & 0.849 & {\bf 0.831} & {\bf 0.817} & {\bf 0.782} & {\bf 0.778}\\\hline
 \end{tabular}}
\end{table*}

\begin{table}[!h]
\centering
\caption{Performance of various overlapping community detection algorithms in terms of three validation measures to detect ground-truth
communities from LFR networks.}\label{result_lfr}
\scalebox{0.8}{
 \begin{tabular}{|c||c|c|c|}
 \hline
\multirow{2}{*}{\bf Algorithms} & \multicolumn{3}{c|}{{\bf LFR}}  \\\cline{2-4}
   & {\bf ONMI} & {\bf $\Omega$ Index} & {\bf F-Score} \\\hline
OSLOM & 0.237 & 0.903 & 0.086  \\\hline
COPRA & 0.974 & 0.976 & 0.731  \\\hline
SLPA  & 0.987 & 0.978 & 0.945  \\\hline
MOSES & 0.852 & 0.972 & 0.803  \\\hline
EAGLE & 0.149 & 0.913 & 0.108  \\\hline
BIGCLAM & 0.908 & 0.851 & 0.785  \\\hline
MaxGenPerm & {\bf 0.984} & {\bf 0.978} & {\bf 0.968}  \\\hline
\end{tabular}}
\end{table}

\section{Comparing MaxGenPerm with Other Baseline Algorithms} 
In this section, we show the performance of all the algorithms in terms of three validation measures (ONMI, Omega Index and F-Score) on
various synthetic and real-world networks. As mentioned in the main text, we run each algorithm on 500 different samples of each
real-world network and the average performance is reported in Table \ref{result_real}. However, for LFR network we retain the original size
with the following parameters: $n$=1000, $\mu$=0.2, $O_n$=5\% and $O_m$=4; and the results are reported in Table \ref{result_lfr}. As we
can observe in both the tables, for most of the cases, MaxGenPerm outperforms other competing algorithms.

\section{ Resolution Limit in Overlapping Community}

In this section, we theoretically prove that using maximizing GenPerm, one can reduce the effect of resolution limit in
overlapping community detection.

\section*{Proof of Theorem 1}
Let us assume that there are three cliques $X$, $Y$ and $Z$ (shown in Figure \ref{theorem}(c)) of size $n_x$, $n_y$, and $n_z$
respectively, such that
$X$ and $Z$ are not connected to each other.
$Y$ is connected to $X$ and $Z$ by two edges ($u_x, v_x$)  and ($u_z, v_z$) respectively, where $u_x \in X$, $u_z \in Z$ and $v_x, v_z \in
Y$. There can be three possible ways mentioned below by which
one can group nodes into different overlapping communities.\\

\noindent {\bf Case 1.} The overlapping community structure shown in Figure \ref{theorem}(d). In this case, one can compute
$P_{g}$ of the network. Note that all the vertices in the network except $u_x$, $v_x$, $u_z$ and $v_z$ are unaffected for any kind of
community assignment and they have $P_{g}$ as 1.

Vertex $u_x$ has no external connection in $C_1$ and $C_2$. In that case, we assume $E_{max}=1$ to avoid the ``divide by zero'' case  for
computing $P_{g}$. \\
{\scriptsize
$P_{g}^{C_1}(u_x)=\frac{n_x-1+1/2}{n_x} - \Big(1-\frac{\frac{(n_x-1)(n_x-2)}{2}}{\frac{n_x(n_x-1)}{2}}\Big) \cdot \frac{(n_x-1+1/2)}{n_x} =
\frac{(n_x-2)(2n_x-1)}{2n_x^{2}}$, \\
$P_{g}^{C_2}(u_x)=\frac{1/2}{n_x}-(1-0)\cdot\frac{1/2}{n_x}=0$.} \\Therefore,
{\scriptsize
\begin{equation}\label{case1.1}
P_{g}(u_x)=P_{g}^{C_1}(u_x)+P_{g}^{C_2}(u_x)=\frac{(n_x-2)(2n_x-1)}{2n_x^{2}} 
\end{equation}}

Similarly for $v_x$, there is no external connection in any of its assigned communities $C_1$, $C_2$ and $C_3$.
{\scriptsize$P_{g}^{C_1}(v_x)=\frac{1/2}{n_y}-(1-0)\cdot\frac{1/2}{n_y}=0$;\\ 
$P_{g}^{C_2}(v_x)=\frac{n_y/2}{n_y} - \Big( 1- \frac{\frac{(n_y-1)(n_y-2)}{2}}{\frac{n_y(n_y-1)}{2}}\Big) \cdot
\frac{n_y/2}{n_y}=\frac{n_y-2}{2n_y}$ \\
 $P_{g}^{C_3}(v_x)=  \frac{\frac{n_y-1}{2}}{n_y} - (1-1)\frac{\frac{n_y-1}{2}}{n_y}=\frac{n_y-1}{2n_y}$}.\\
Therefore,
{\scriptsize
\begin{equation}\label{case1.2}
P_{g}(v_x)=\frac{n_y-2}{2n_y}+\frac{n_y-1}{2n_y}=\frac{2n_y-3}{2n_y}
\end{equation}}

\begin{figure*}
\centering
\scalebox{1}{
\includegraphics[width=\textwidth]{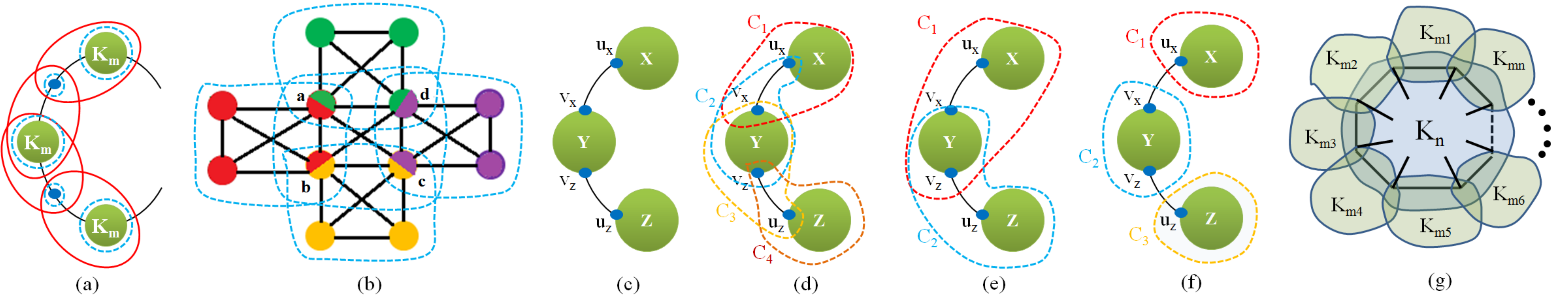}
}
\caption{(Color online) Toy examples used to demonstrate the theorems ($K_x$: clique of size $x$).}
\label{theorem}
\end{figure*}

The connectivity of $v_z$ is similar to $v_x$.\\
{\scriptsize
$P_{g}^{C_2}(v_z)=\frac{\frac{n_y-1}{2}}{n_y}-(1-1)\cdot\frac{\frac{n_y-1}{2}}{n_y}=\frac{n_y-1}{2n_y}$;\\
 $P_{g}^{C_3}(v_z)= \frac{n_y/2}{n_y} - \Big( 1- \frac{\frac{(n_y-1)(n_y-2)}{2}}{\frac{n_y(n_y-1)}{2}}\Big) \cdot
\frac{n_y/2}{n_y}=\frac{n_y-2}{2n_y}$ ;\\
$P_{g}^{C_4}(v_z)=0$}.\\ 
Therefore, 
{\scriptsize
\begin{equation}\label{case1.3}
 P_{g}(v_z)=\frac{n_y-1}{2n_y}+\frac{n_y-2}{2n_y}+0=\frac{2n_y-3}{2n_y}
\end{equation}}

Further, the connectivity of vertex $u_z$ is similar to that of $u_x$ except the degree of $u_z$ being $n_z$.
Therefore,
{\scriptsize
\begin{equation}\label{case1.4}
P_{g}(u_z)=\frac{(n_z-2)(2n_z-1)}{2n_z^{2}}
\end{equation}}

Now combining the GenPerm of the affected vertices shown in Equations \ref{case1.1}, \ref{case1.2}, \ref{case1.3} and
\ref{case1.4}, we obtain:
{\scriptsize
\begin{equation}\label{case1}
\begin{split}
 P_{g}^1=P_{g}(u_x)+P_{g}(v_x)+P_{g}(v_z)+P_{g}(u_z)\\
=\frac{(n_x-2)(2n_x-1)}{2n_x^{2}}+\frac{(n_z-2)(2n_z-1)}{2n_z^{2}}+\frac{2(2n_y-3)}{2n_y}\\
=4-\frac{5}{2n_x}-\frac{5}{2n_z}-\frac{3}{n_y}+\frac{1}{n_x^2}+\frac{1}{n_z^2}
\end{split}
\end{equation}}

\noindent {\bf Case 2.} The overlapping community structure shown in Figure \ref{theorem}(e). We again calculate $P_{g}$ for the affected
vertices as follows.

For vertex $u_x$ which belongs to community $C_1$ only, there is no external connection, and no internal edges are shared. Therefore,
{\scriptsize
\begin{equation}\label{case2.1}
\begin{split}
P_{g}(u_x)=P_{g}^{C_1}(u_x)=\frac{n_x}{n_x} - \Big( 1 - \frac{\frac{(n_x-1)(n_x-2)}{2}}{\frac{n_x(n_x-1)}{2}}\Big) \cdot
\frac{n_x}{n_x}\\
=\frac{n_x-2}{n_x}
\end{split}
\end{equation}}

For vertex $v_x$ belonging to both $C_1$ and $C_2$, the internal edges in clique $Y$ are shared by two communities. Therefore,\\
{\scriptsize
$P_{g}^{C_1}(v_x)=\frac{\frac{n_y-1}{2}+1}{n_y}-(1-\frac{n_y-2}{n_y})\cdot(\frac{\frac{n_y-1}{2}+1}{n_y})=\frac{(n_y+1)(n_y-2)}{2n_y^2}
=\frac{n_y^2-n_y-2}{2n_y^2}$;\\
 $P_{g}^{C_2}(v_x)=  \frac{\frac{n_y-1}{2}}{n_y} - (1 - 1) \cdot \frac{\frac{n_y-1}{2}}{n_y}=\frac{n_y-1}{2n_y}$}.\\
Therefore,
{\scriptsize
\begin{equation}\label{case2.2}
 P_{g}(v_x)=P_{g}^{C_1}(v_x)+P_{g}^{C_2}(v_x)=\frac{n_y^2-n_y-1}{n_y^2}
\end{equation}}

Similarly, the GenPerm of $v_z$ is similar to that of $v_x$. Therefore,\\
{\scriptsize
\begin{equation}\label{case2.3}
P_{g}(v_z)=\frac{n_y^2-n_y-1}{n_y^2}
\end{equation}}

The connectivity of $u_z$ is similar to that of $u_x$, except the degree of $u_z$ being $n_z$. Therefore following Equation \ref{case2.1},
{\scriptsize
\begin{equation}\label{case2.4}
P_{g}(u_z)=\frac{n_z-2}{n_z}
\end{equation}}

Now combining the GenPerm of all the affected vertices, we obtain:
{\scriptsize
\begin{equation}\label{case2}
\begin{split}
 P_{g}^2=P_{g}(u_x)+P_{g}(v_x)+P_{g}(v_z)+P_{g}(u_z)\\
 =\frac{n_x-2}{n_x}+\frac{n_z-2}{n_z}+\frac{2(n_y^2-n_y-1)}{n_y^2}\\
 =4-\frac{2}{n_x}-\frac{2}{n_y}-\frac{2}{n_z}-\frac{2}{n_y^2}
\end{split}
\end{equation}}

\noindent {\bf Case 3.} The  community structure shown in Figure \ref{theorem}(f). Note that in this case, each of $u_x$,
$v_x$, $v_z$ and $u_z$ have one external neighbor and the internal clustering coefficient is 1. Therefore,
{\scriptsize
$P_{g}(u_x)=P_{g}^{C_1}(u_x)=\frac{n_x-1}{n_x}-(1-1)\cdot\frac{n_x-1}{n_x}=\frac{n_x-1}{n_x}$.}
Similarly, {\scriptsize$P_{g}(v_x)= \frac{n_y-1}{n_y}$, $P_{g}(v_z)=\frac{n_y-1}{n_y}$ and $P_{g}(u_z)=\frac{n_z-1}{n_z}$.}

Therefore, combining the GenPerm of the affected vertices, we obtain:
{\scriptsize
\begin{equation}\label{case3}
\begin{split}
 P_{g}^3=P_{g}(u_x)+P_{g}(v_x)+P_{g}(v_z)+P_{g}(u_z)\\
 =\frac{n_x-1}{n_x}+\frac{2(n_y-1)}{n_y}+\frac{n_z-1}{n_z}\\
=4-\frac{1}{n_x}-\frac{2}{n_y}-\frac{1}{n_z}
\end{split}
\end{equation}}

\noindent{\bf Case 1 vs. Case 3:}
Subtracting Equation \ref{case3} from Equation \ref{case1}, we obtain
{\scriptsize
\begin{equation}
\begin{split}
  P_{g}^1 - P_{g}^3 = -\frac{3}{2n_x}-\frac{3}{2n_z}-\frac{1}{n_y}+\frac{1}{n_x^2}+\frac{1}{n_z^2}
\end{split}
\end{equation}}
Now, {\scriptsize$P_{g}^1 - P_{g}^3<0$}, if {\scriptsize$\frac{3}{2n_x}+\frac{3}{2n_z}+\frac{1}{n_y}>\frac{1}{n_x^2}+\frac{1}{n_z^2}$}. Note
that
$\frac{3}{2n_x}>\frac{1}{n_x^2}$, $\frac{3}{2n_z}>\frac{1}{n_z^2}$. So the above condition always holds. Therefore, GenPerm
will always prefer case 3 in comparison to case 1 during maximization.\\

\noindent{\bf Case 2 vs. Case 3:}
Subtracting Equation \ref{case3} from Equation \ref{case2}, we obtain
{\scriptsize
\begin{equation}
\begin{split}
  P_{g}^2 - P_{g}^3 = -\frac{1}{n_x}-\frac{1}{n_z}-\frac{4}{n_y^2} < 0
\end{split}
\end{equation} }

Therefore, maximum value of $P_{g}$ will always be obtained for case 3 in comparison to case 2.

From this two comparison, we observe that the highest $P_{g}$ is obtained when $X$, $Y$ and $Z$ are three separate communities as shown in
Figure \ref{theorem}(f). Note that there can be other scenarios as well such as {\bf Case 4} where any of the two consecutive cliques are
combined together to form a community along with the remaining clique as separate community, thus forming two non-overlapping communities,
or {\bf Case 5} where all three cliques are combined to form one single community. In case of permanence \cite{chakraborty_kdd}, the author
showed
that if one can maximize permanence, case 3 always wins
with respect to cases 4 and 5. Therefore, we can conclude that for the network shown in Figure \ref{theorem}(c), maximizing $P_{g}$ always
yields the community structure shown in Figure \ref{theorem}(f), thus reducing the effect of resolution limit in community detection
irrespective of the community size.

\section*{Proof of Theorem 2}
Let us assume the toy example shown in Figure \ref{theorem}(b), a generalized version of which is shown in Figure \ref{theorem}(g). Assume
that there is a central clique $X$ of size $n$ which is surrounded
by $n$ cliques, $C_1$, $C_2$, ...., $C_n$ with size $n_1$, $n_2$,..., $n_n$ respectively such that each of the surrounding cliques share one
edge with
$X$.

Now let us assume the community structure as shown by the broken lines in Figure \ref{theorem}(b) where there are $n+1$
communities, namely $X$, $C_1$, $C_2$, ...., $C_n$, and the nodes in $X$ such as $a$, $b$, $c$, $d$, ... are overlapping nodes and each
of them belong to two
communities (we call these overlapping nodes together as set $V_{ov}$). One can measure the GenPerm of the
graph in this setting. It is simple to show that the vertices except $V_{ov}$ have $P_{g}$ as 1.  However, we calculate $P_{g}$ of each
vertex in $V_{ov}$ as follows. Note that each node in $V_{ov}$ has equal $P_{g}$ value.

Let us consider vertex $a$ ($\in V_{ov}$) which shares $C_1$, $C_2$ and $X$ communities. The degree $a$ is $D(a)=n_1+n_2+n-5$. In
community $C_1$, $a$ has $n_1-2$ internal connections which
are not shared, but edge ($a,d$) is shared with both $C_1$ and $X$. It has no external connection and its internal clustering coefficient in
this community is 1. Therefore,
{\scriptsize
\begin{equation}\label{theorem2.1}
\begin{split}
 P_{g}^{C_1}(a)=\frac{n_1-2+1/2}{n_1+n_2+n-5} - (1-1)\cdot(\frac{n_1-2+1/2}{n_1+n_2+n-5})\\
=\frac{n_1-3/2}{n_1+n_2+n-5}
\end{split}
\end{equation}}

Similarly, the GenPerm of $a$ in community $C_2$ is as follows.
{\scriptsize
\begin{equation}\label{theorem2.2}
\begin{split}
 P_{g}^{C_2}(a)=\frac{n_2-3/2}{n_1+n_2+n-5}
\end{split}
\end{equation}}

In community $X$, $a$ has $n-1$ internal connections among which two edges, namely ($a,b$) and ($a,d$) are part of two communities.
It has no external connection and the internal clustering coefficient is 1.
Therefore,
{\scriptsize
\begin{equation}\label{theorem2.3}
 \begin{split}
  P_{g}^X(a)=\frac{n-3+1/2+1/2}{n_1+n_2+n-5}\\
 = \frac{n-2}{n_1+n_2+n-5}
 \end{split}
\end{equation}}

Combining Equations \ref{theorem2.1}, \ref{theorem2.2} and \ref{theorem2.3}, we obtain the overall GenPerm of vertex $a$ as
follows:
{\scriptsize
\begin{equation}
\begin{split}
P_{g}(a)= P_{g}^{C_1}(a) + P_{g}^{C_2}(a) + P_{g}^X(a)\\
=\frac{n_1+n_2+n-5}{n_1+n_2+n-5}=1
\end{split}
\end{equation}}

Similarly, one can show that the GenPerm of each vertex in $V_{ov}$ is 1. Since the maximum value of $P_{g}$ for a vertex
is 1, this configuration is the best community assignment one can obtain using maximizing $P_{g}$.


\end{document}